\documentclass[11pt, twoside]{article}
\usepackage{amsmath}
\usepackage{amssymb}			
\usepackage{amsthm}
\usepackage{amsfonts}
\usepackage{mathtools}
\usepackage{empheq}
\usepackage[mathscr]{eucal}

\usepackage{tikz}
\usetikzlibrary{tikzmark,fit}							

\usepackage{lmodern} 											
\usepackage[T1]{fontenc} 											
\usepackage{microtype} 											
\usepackage[lmargin=20mm, hmarginratio=1:1, top=25mm, bottom=25mm, columnsep=20pt]{geometry} 		
\usepackage{multicol} 											

\usepackage{abstract} 									

\newtheoremstyle{def}		
	{2.5\topsep}									
	{5pt}											
	{\upshape}									
	{}											
	{\bfseries}										
	{}											
	{\newline}										
	{\underline{\thmname{#1}}:\ \thmnote{ #3}}			

\newtheoremstyle{clm}{2.5\topsep}{-\topsep}{\upshape}{}{\bfseries}{:}{7pt}	{\thmname{#1}\thmnumber{ #2}}	

\newtheoremstyle{th}	{2.5\topsep}{-\topsep}{\itshape}{}{\bfseries}{}{\newline}{\underline{\thmname{#1}\thmnumber{ #2}}}
	
\newtheoremstyle{lmm}{2.5\topsep}{-\topsep}{\upshape}{}{\itshape}{:}{7pt}{\thmname{#1}\thmnumber{ #2}}

\theoremstyle{th}

\theoremstyle{clm}

\theoremstyle{def}

\theoremstyle{lmm}

\usepackage{fancyhdr} 			

\usepackage[titletoc]{appendix}			

\usepackage{cite}
\usepackage{paralist} 										
\usepackage[hang, small,labelfont=bf,up,textfont=it,up]{caption} 		
\usepackage{booktabs} 										
\usepackage{float} 					
\usepackage{cases}
\usepackage{multicol}

\usepackage{slashed}

\usepackage{hyperref} 				

\usepackage{tensind}			
\tensordelimiter{?}

\title{\vspace{15mm}\fontsize{14pt}{15pt}\selectfont\textbf{Black Holes in Higher Dimensional General Relativity}\vspace{20mm}} 

\author{
	\large
	\textsc{David Pere\~niguez Rodriguez}\\[12mm] 			
	\textsc{Under the supervision of}\\
	\textsc{Dr. Jorge Santos}
}
\date{}

\begin{document}				

\begin{titlepage} 			
\centering

\maketitle 						
\thispagestyle{fancy}
\fancyfoot{}
\fancyhead[C]{University of Cambridge. Mathematical Tripos - Part III Essay. 2017-2018} 

\vspace{20mm}

\begin{abstract}

\noindent In this essay some aspects of General Relativity in higher dimensions are reviewed. The work presented draws a path within the wide landscape of higher dimensional black holes theory towards a specific objective: understanding how the combination of extended black objects and ultra-spinning rotation dynamics give rise to black hole topologies that depart from sphericity. The discussion aims to offer a compromise between giving a reasonably wide picture of the area of higher dimensional black holes, while at the same time going into detailed calculations that make manifest the enhanced nature of General Relativity in $d>4$. 

\end{abstract}

\vfill

\date{\large\today}

\end{titlepage}


\pagestyle{empty}
\newpage
\tableofcontents

\newpage


\setcounter{page}{1}
\pagestyle{fancy}
\fancyhead{}
\fancyfoot{}
\fancyfoot[C]{-- ~\thepage~ --} 		

\renewcommand{\headrulewidth}{0pt}

\section{Introduction}
\subsection{Motivation}
The study of General Relativity (GR) in higher dimensions has been a topic of intense research activity in the recent years. One of the reasons why this is so is that, in the current paradigm of theoretical physics, the study of Einstein's theory of Gravitation in arbitrary $d$ has a wide variety of applications. In the following lines we cite and comment some of the most relevant ones. However, it is almost a century ago that the idea of considering extra dimensions in GR started to pop into the intuition of theoretical physicists. In the 1920's, Kaluza and Klein tried to unify gravity and electromagnetism via the dimensional reduction of five dimensional, pure-gravity GR with a compact extra dimension in the topology of the spacetime, namely $S^{1}$ \cite{KK1,KK2}. Although the theory turned out not to provide a full unification, GR and specially its black hole solutions in spacetimes with compact extra dimensions are still a topic of interest in theoretical physics. Nowadays, some of the most relevant applications of higher dimensional GR are found in the following areas
\begin{itemize}
\item String Theory: Considered by many as a strong candidate for a fundamental theory of gravitation, String Theory requires several extra dimensions. Furthermore, some of the most relevant achievements towards the microstate description of black holes have been obtained in dimension larger than four, see e.g. the work done by Strominger and Vafa in \cite{BHentropy} in which the Bekenstein-Hawking entropy law was recovered from a microscopic string theoretic description.
\item Brane World models: It has been suggested \cite{Braneworld1} that our four dimensional universe is actually a membrane embedded in a five dimensional, $\mathbb{Z}_{2}$-symmetric spacetime. While the Standard Model fields are trapped in the membrane, gravity can access the bulk and fluctuate around the membrane. Consequently, understanding GR in higher dimensions becomes a crucial issue when constructing such braneworld models.
\item AdS/CFT: Or more generally the gauge gravity dualities \cite{Malda} provide, roughly speaking, a dictionary between classical black holes in $d$ dimensions and QFT's in $d-1$ dimensions without gravity. Working in the gravity side of the duality, several achievements have been carried out towards the understanding of aspects of QCD \cite{Mateos} and Condensed Matter Physics \cite{Santos}, among other results. Furthermore, some extensions of the original AdS/CFT correspondence have been conjectured in which classical $d-1$ dimensional black holes living on the brane of the $AdS_{d}$ brane world are dual to black hole solutions of the 1-loop-corrected Einstein equations \cite{Conjecture1}. Hence, a better understanding of GR in, say $d=5$, can lead to the understanding of the (effective) coupling between the gravitational field and quantum matter in our $4$-dimensional world.     
\end{itemize}
Although this applications provide more than enough reasons to work on the understanding of higher dimensional GR, its study is also of intrinsic interest: in the heart of the interests of any theoretical physicist there is always the desire of studying, probing and understanding a theory in its full glory regardless of how close it is from resembling the real world. It is this general study what ends up providing a precise definition of the sector of the theory that can describe the physical world, and what gives a deep insight in the physics of such sector.
\subsection{General Relativity in arbitrary dimensions}
It turns out that GR is remarkably sensitive to the spacetime dimension, $d$. Here we shall take a brief tour from $d=2$ to $d\to\infty$ discussing how pure gravity (i.e. with no matter present and with $\Lambda=0$) behaves in each case according to GR.

If the purpose of GR is to describe the geometry of spacetime, then one of the minimal expressions of the theory is its formulation in the two dimensional case in which there are only one 'time direction' and one 'spatial direction'. Nevertheless, in such case the Gauss-Bonnet theorem tells us that the Einstein-Hilbert action is a topological invariant: for a given spacetime topology, the vacuum field equations will be identically satisfied for \textit{any} metric tensor. The allowed gravitational fields are filtered out by the spacetime topology rather than a dynamical equation, so the study of GR pure gravity in $d=2$ is not of much interest\footnote{However, it is a completely different story when $\Lambda\ne0$ and \textit{quantum} matter is considered. See, e.g. \cite{BirrellDavis}.}. In $d=3$ the Einstein-Hilbert action is no more a topological invariant, but the Weyl tensor is identically zero \cite{WaldGR}. Then so is the Riemann tensor in the absence of matter and cosmological constant. It is not until $d=4$ that the gravitational field can fluctuate in the strict vacuum. However, the so-called \textit{no hair theorems} tell us that the most fundamental objects of the theory, stationary black holes, are fully determined by their mass and angular momenta (see, any of \cite{WaldGR,LargeScale,Carroll}). In $d=4$, Hawking's theorems \cite{Hawking} on the horizon topology and \textit{horizon rigidity} show respectively that $i)$ black holes must be topological 2-spheres, and that $ii)$ stationary black holes are also axisymmetric. Furthermore, there are very strong arguments in favour of the dynamical stability of black holes in $d=4$. As we will discover throughout this essay, many of such features do not hold in $d>4$. Actually, the fact that four dimensional black holes are so constrained is a very special feature of $d=4$. In particular, we will see that in $d>4$ the horizon topology constraints are less restrictive, allowing for topologies such as rings $S^{1}\times S^{d-3}$ or even with different connected components (stationary multi-black hole solutions). Regarding black hole uniqueness, we will see in detail that it is explicitly broken in $d=5$ and there are several reasons to expect a large degeneracy of black hole solutions in $d\geq6$. In addition, the fact that now the horizon admits dimensions that are "non-equally large" (for instance, a black hole $S^{1}\times S^{2}$ for which the length scale of the $S^{1}$ is much larger than that of the $S^{2}$) turns out to result in a dynamical instability. Finally, if one considers the large $d$ limit of GR it turns out that the theory becomes extremely simplified: black holes behave as non-interacting particles that do not absorb or emit radiation of \textit{finite} frequency \cite{LargeD}.

The clear enhancement of features exhibited by solutions of higher dimensional GR can be argued to be a consequence of three phenomena: 
\begin{itemize}
\item Extended black objects: As we will see in the following section, in $d>4$ GR admits black holes with horizons with non compact spatial dimensions, or with some directions way larger than the others.
\item New dynamics of rotation: In $d=4$, stationary black holes can not rotate arbitrarily fast. Exceeding certain bound in their speed of rotation results in a naked singularity. This is not the case when $d\geq5$: black holes will be allowed to rotate arbitrarily fast and in different rotation planes simultaneously. Actually, this difference on the dynamics of rotation could have been expected. For simplicity, consider a delta function distribution of matter with support at the origin in $d$ spacetime dimensions. The associated Newtonian potential goes like
\begin{equation}\label{eq:0001}
-\frac{GM}{r^{d-3}}.
\end{equation}
At the same time, since in any dimension rotation is defined within a plane, the so-called centrifugal barrier goes like
\begin{equation}\label{eq:0002}
\frac{J^{2}}{M^{2}r^{2}}
\end{equation}
and has no explicit dependence in $d$. It is manifest that the relation between this two terms will vary significantly with $d$. Therefore, since the geodesic generators of the black hole horizons could be orbits followed by massless particles, one expects the dimensional dependence of the balance between the collapse and the centrifugal force to be also significant for black holes. 
\item Instability of extended black objects: it has been shown that extended black objects are unstable \cite{GL}. Then, when one of the directions of the horizon becomes much larger than the others this instability enters into the game pushing the black hole towards more stable configurations.
\end{itemize}
The combination of the first two points allows one to (heuristically speaking) construct new black hole shapes with contractible directions, and compensate the gravitational collapse via fast rotation. For instance, it will be possible to construct a "donut black hole", and compensate the collapse of the contractible circles through the centrifugal force originated by rotation in the plane of the donut. Although this way of thinking might sound too simplistic, we will see that these heuristic arguments turn out to be very accurate and sometimes even exact\footnote{The remarkable difference, however, is that now it is the donut that would eat us, and not the other way around.}. 
\subsection{Outline}
The content of this work draws a curve within the wide landscape of higher dimensional GR towards the achievement of a main objective: understanding how the combination of extended black objects and the new rotation dynamics leads to the existence of black holes with new shapes and that do not have analogue in $d=4$. We will restrict to the vacuum with no cosmological constant. In section 2 we introduce the basic concepts that will be needed throughout the essay: we briefly construct the simplest solutions of extended black objects, the homogeneous black $p$-branes, and define the charges associated to an asymptotically flat solution. Section 3 is devoted to study the non-upper-bounded rotation dynamics exhibited by the natural extension of the Kerr solution to higher dimensions: the Myers-Perry black hole. The detailed calculations are performed in the single-spin case, and the discussion is prolongated also to the general case of several spins. Section 4 closes our study presenting the family of black rings with rotation along the plane of the ring. Among other aspects, we study the thin ring mechanics and its resemblance to Newtonian strings, and the connection between 'plump' black rings to the single-spin MP black hole in $d=5$.

 Finally, we want to emphasise that, unfortunately, very interesting topics such as stability against gravitational perturbations, solution generating techniques, the algebraic classification of solutions or the effective blackfold theory (the latter being probably the analytic continuation of this essay) will only be mentioned or not even present at all in this study, due to both timelike and spacelike constraints in the realisation of the essay.
\section{Basic Concepts: Conserved Charges and Extended Black Objects}
\subsection{Conserved Charges}
 In the vacuum with no cosmological constant, GR is extended to higher dimensional spacetimes in the simplest way\footnote{Throughout this work we will use the same units and numerical factors as in \cite{EmparanReallReview}},
\begin{equation}\label{eq:00001}
I=\frac{1}{16\pi G}\int d^{d}x\sqrt{-g}R.
\end{equation}
We will follow \cite{MPSolution} to define the conserved charges of an asymptotically flat solution. Instead of considering the Komar charges from their definition, we will identify the charges by inspection of the asymptotic field of the solution. The idea is first to write down the far field created by matter present in a compact region around the origin. Then, comparing this field to the asymptotic field of a solution, the charges can be easily identified and related to the parameters of the solution.

Consider an energy momentum tensor $T_{\mu\nu}$ compactly supported around the origin. The linearised Einstein equations read \cite{WaldGR,Carroll}
\begin{equation}\label{eq:00002}
\partial^{\rho}\partial_{\rho}\bar{h}_{\mu\nu}=-16\pi G T_{\mu\nu}.
\end{equation}
where $\bar{h}_{\mu\nu}=h_{\mu\nu}-\frac{1}{2}h\eta_{\mu\nu}$ and we have imposed the Lorentz gauge $\partial^{\nu}\bar{h}_{\mu\nu}=0$. Instead of solving \eqref{eq:00002} generally (see, e.g. \cite{WaldGR}) and then approximate the field far from the source, we can follow a more straightforward procedure. The field created \textit{far from the source} by matter in a compact region around the origin is the same as the field created \textit{everywhere} by a new energy momentum tensor $\tilde{T}_{\mu\nu}$,
\begin{align}\label{eq:00003}
\tilde{T}_{00}=M\delta(\bold{x})+\frac{1}{2}I_{ij}\partial_{i}\partial_{j}\delta(\bold{x})\\ \nonumber
\tilde{T}_{0i}=\frac{1}{2}\left(\dot{I}_{ij}+J_{ij}\right)\partial_{j}\delta(\bold{x})\\ \nonumber
\tilde{T}_{ij}=\frac{1}{2}\ddot{I}_{ij}\delta(\bold{x})
\end{align}
if we identify the constants $M$ and $J_{ij}=J_{[ij]}$ with the physical energy momentum tensor as
\begin{equation}\label{eq:00004}
M=\int d^{d-1}x T_{00},\,\,\,\,\,\,\, J_{ij}=-2\int d^{d-1}x T_{0[i}x_{j]}
\end{equation}
and $I_{ij}(t)$ with the second moment of the physical energy density
\begin{equation}\label{eq:00005}
I_{ij}=\int T_{00}x^{i}x^{j}d^{d-1}x.
\end{equation}
A particularly neat justification of the use of $\tilde{T}_{\mu\nu}$ is given in appendix A of \cite{Cosmo}. In our case it will be enough to restrict to $\tilde{T}_{\mu\nu}$ with $I_{ij}=0$, which corresponds to matter localised in a very small region at the origin. Plugging such $\tilde{T}_{\mu\nu}$ in \eqref{eq:00002}, one can easily Fourier-transform at both sides (or use the fundamental solution of the problem $\square f=\rho$) and invert the definition of $\bar{h}_{\mu\nu}$ to get
\begin{align}\label{eq:00006}
ds^{2}=-\left(1-\frac{16\pi G}{(d-2)\Omega_{d-2}}\frac{M}{r^{d-3}}\right)dt^{2}-\frac{8\pi G}{\Omega_{d-2}}\frac{x^{k}J_{ki}}{r^{d-1}}dtdx^{i}+\\ \nonumber
+\left(1+\frac{16\pi G}{(d-2)(d-3)}\frac{M}{\Omega_{d-2}}\frac{1}{r^{d-3}}\right)\delta_{ij}dx^{i}dx^{j}.
\end{align}
Without loss of generality, we can choose the spatial coordinates $x^{i}$ such that $J_{ij}$ takes a block diagonal form, with $N:=\lfloor\frac{d-1}{2}\rfloor$ block matrices 
\begin{equation}\label{eq:00007}
J_{a}\left(\begin{array}{cc}0&1\\
-1&0\end{array}\right)\,\,\,\,\,\,\,\,\,\, J_{a}\in\mathbb{R},\,\,\,\,\,\,\, a=1,...,N
\end{equation}
plus a row and a column of zeros in the case of even $d$. The interpretation is clear: each $J_{a}$ is the angular momenta due to rotation in a plane $(x_{2a-1},x_{2a})$. Organising the spatial coordinates in such planes $(x_{2a-1},x_{2a})$ and introducing the usual plane polar coordinates $x_{2a-1}=r_{a}\cos\phi_{a}$, $x_{2a}=r_{a}\sin\phi_{a}$ in each of them, one gets (suspending momentarily the summation convention on $a$)
\begin{align}\label{eq:00008}
ds^{2}=-\left(1-\frac{16\pi G}{(d-2)\Omega_{d-2}}\frac{M}{r^{d-3}}\right)dt^{2}-\frac{8\pi G}{\Omega_{d-2}}\sum_{a}J_{a}\left(\frac{r_{a}}{r}\right)^{2}\frac{d\phi_{a}dt}{r^{d-3}}+\\ \nonumber
+\left(1+\frac{16\pi G}{(d-2)(d-3)\Omega_{d-2}}\frac{M}{r^{d-3}}\right)\sum_{a}(dr_{a}^{2}+r_{a}^{2}d\phi_{a}^{2}).
\end{align}
Let us now introduce another set of coordinates that will be convenient in several points throughout this work. Consider the \textit{direction cosines} $\mu_{a}:=r_{a}/r$. These satisfy $\mu_{a}^{2}=1$ and $\mu_{a}^{2}+z^{2}/r^{2}=1$ for $d$ odd and even, respectively\footnote{Notice that, because of such constraint, $(t,\mu_{a})$ is \textit{not} a local chart. However, eliminating one direction cosine and using instead, say $r$, one gets a well defined local chart $(t,r,\mu_{1},...,\mu_{N-1})$, and similarly for the case $d$ even. For convenience, however, we will typically write the metric in terms of all the direction cosines.}. Using them the metric reads
\begin{align}\label{eq:00009}
ds^{2}=-\left(1-\frac{16\pi G}{(d-2)\Omega_{d-2}}\frac{M}{r^{d-3}}\right)dt^{2}-\frac{8\pi G}{\Omega_{d-2}}\sum_{a}J_{a}\mu_{a}^{2}\frac{d\phi_{a}dt}{r^{d-3}}+\\ \nonumber
+\left(1+\frac{16\pi G}{(d-2)(d-3)\Omega_{d-2}}\frac{M}{r^{d-3}}\right)\left(dr^{2}+r^{2}d\alpha^{2}+r^{2}\sum_{a}(d\mu_{a}^{2}+\mu_{a}^{2}d\phi_{a}^{2})\right).
\end{align}
where $\alpha=z/r$ in the case of even $d$, and $\alpha=0$ for odd $d$. For a given asymptotically flat solution, we can write the field in the asymptotical region and read off the mass and angular momenta via comparison with any of \eqref{eq:00006}, \eqref{eq:00008} or \eqref{eq:00009}.

Finally, in order to perform meaningful comparisons between different black hole solutions it is convenient to define appropriate dimensionless magnitudes. For that, we need to fix a scale where the solutions live. The Riemann tensor in the vacuum is (equal to the Weyl tensor so it is) conformally invariant; the scale must be set by a physical parameter of the solution. The most natural choice is to use the mass $M$ of the black hole, as defined in \eqref{eq:00006}. Hence, we can follow \cite{BRMatchedAsym} and define a dimensionless area $a_{H}$ and spins $j_{a}$ as
\begin{equation}\label{eq:000010}
j_{a}^{d-3}=\frac{\Omega_{d-3}(d-2)^{d-2}}{2^{d+1}(d-3)^{\frac{d-3}{2}}}\frac{J_{a}^{d-3}}{GM^{d-2}},\,\,\,\,\,\,\,\,\, a_{H}^{d-3}=\frac{\Omega_{d-3}}{2(16\pi)^{d-3}}(d-2)^{d-2}\left(\frac{d-4}{d-3}\right)^{\frac{d-3}{2}}\frac{\mathcal{A}_{H}^{d-3}}{(GM)^{d-2}},
\end{equation}
where $\Omega_{d}$ is the area of a unit $d$-sphere. Then, for a given mass and dimension we can compare in a meaningful way, for example, the relation $a_{H}(j_{a})$ of different solutions.  
\subsection{Schwarzschild-Tangherlini and Black $p$-branes}
Here we obtain two basic solutions that will be relevant in several discussions throughout the essay. Following the chronological order of the discoveries of four dimensional GR, we can start by generalising the Schwarzschild solution to an arbitrary number of dimensions.  We want it to be static and spherically symmetric, so we can try with the form
\begin{equation}\label{eq:000011}
ds^{2}=-f(r)dt^{2}+\frac{dr^{2}}{f(r)}+r^{2}d\Omega_{d-2}^{2}.
\end{equation}
The interesting observation now is that in $d$ dimensions the Newtonian potential $\sim h_{00}$ goes like \eqref{eq:0001}. It turns out that the naive substitution $1/r\rightarrow1/r^{d-3}$ in the function $f(r)$ of the four dimensional Schwarzschild solution provides the correct answer \cite{dSchwarzschildOriginal}
\begin{equation}\label{eq:000012}
ds^{2}=-\left(1-\frac{\mu}{r^{d-3}}\right)dt^{2}+\frac{dr^{2}}{\left(1-\frac{\mu}{r^{d-3}}\right)}+r^{2}d\Omega_{d-2}^{2}.
\end{equation}
The mass of this solution can be immediately read off 
\begin{equation}\label{eq:000013}
\mu=\frac{16\pi GM}{(d-2)\Omega_{d-2}}.
\end{equation}
Now that we have a solution of the vacuum Einstein equations in arbitrary $d$, it is very simple to obtain a solution in $d+p$ dimensions. Let $(\mathcal{B},g_{d})$ be a vacuum black hole solution in $d$ dimensions, and let $(\mathbb{R}^{p},\delta)$ be the euclidean space in $p$ dimensions. The $(d+p)$-dimensional manifold $\mathcal{B}\times\mathbb{R}^{p}$ admits the metric
\begin{equation}\label{eq:000014}
g_{d+p}=\pi^{*}_{\mathcal{B}}g_{d}+\pi^{*}_{\mathbb{R}^{p}}\delta,
\end{equation}
being $\pi_{\mathcal{B}}$ and $\pi_{\mathbb{R}^{p}}$ the respective projections. In coordinates and in terms of line elements, we have
\begin{equation}\label{eq:000015}
ds^{2}_{d+p}=ds^{2}_{d}(\mathcal{B})+\delta_{ij}dx^{i}dx^{j}.
\end{equation}
Since both $g_{d}$ and $\delta$ are Ricci-flat (meaning $R_{ab}=0$), so is $g_{d+p}$ and consequently constitutes a vacuum solution of GR in $d+p$ dimensions. This family of solutions are the homogeneous \textit{black $p$-branes}. However, in the case $p=1$, \eqref{eq:000015} is referred to as a \textit{black string}. Notice that, if the topology of (the spatial cross sections of) the horizon of $\mathcal{B}$ is $\Sigma_{H}$, then the associated homogeneous black $p$-brane has horizon topology $\Sigma_{H}\times \mathbb{R}^{p}$ so it constitutes an \textit{extended black object}. Since we already have a black hole solution in any $d\geq4$ \eqref{eq:000012}, black $p$-branes and consequently extended black objects exist in any dimension $d+p>4$. This fact is crucial for the development of the forthcoming discussions: black $p$-branes can be boosted, bent, rippled and even fragmented so they provide a very versatile object with which one can guess and approximate how a black hole of a certain shape might look like. In our case we will only bend and boost black strings in order to construct the so-called five dimensional black ring solutions.
\subsection{The Gregory-Laflamme instability}
 Although this essay will not deal with the study of the dynamic stability of black hole solutions, there is a result that will be very relevant for our work and that we have already mentioned: when one direction of the horizon becomes way larger than the others, then the black hole may become unstable against gravitational perturbations. This kind of instabilities are commonly referred to as Gregory-Laflamme (GL) instabilities \cite{GL}. Lets us give a glimpse of how these work.

One of the simplest cases in which one direction of the horizon is way larger than the others is the black string constructed by adding a flat (and for the moment non compact) direction to \eqref{eq:000012}, whit horizon radius $r_{H}$. Although \eqref{eq:000012} itself has been shown to be stable against linearised gravitational perturbations \cite{SchStable}, in the black string the modes can now also propagate along the extended direction $z$, so that roughly speaking they go like $\sim e^{-i(\Omega t-kz)}$. Similarly to any process of wave propagation, the equations of motion establish certain dispersion relation $\Omega(k)$. It turns out that, according to such dispersion relation, when the wave length of the propagation is larger than the Schwarzschild radius $k<k_{GL}\sim2\pi/r_{H}$, then $\text{Im}(\Omega)>0$. This means that such modes go like $\sim e^{\text{Im}(\Omega)t}$ so they constitute an instability because they grow with time. One of the possible end points of such instability is that the string could end up fragmenting to give rise to localised black holes. In particular, notice that the GL instability implies that circular black strings with radius $R>r_{H}/2\pi$ are unstable, as the circle is large enough to accommodate the unstable modes. In such cases, a fragmentation of the string would produce localised Kaluza-Klein black holes. As a last remark, one finds that for $k=k_{GL}$ the perturbations are \textit{static}, i.e. do not exhibit time dependence. Since these perturbations have nontrivial $z$ dependence, they give rise to a rippled black string. Such black strings are said to be \textit{inhomogeneous}. In the case of black $p$-branes the same results hold for perturbations propagating in an arbitrary direction $\bold{k}$ along the (i.e. tangent to) the $p$ flat extra dimensions.

From all this discussion, what will end up being relevant for us is the fact that, when a black hole gets deformed so that one of the directions is much larger than the others, perturbations can propagate along the extended directions and GL instabilities pop into the game.   
\section{New rotation dynamics: The Myers-Perry Black Holes}
The presence of extra dimensions allow for a richer set of configurations in the rotation regimes of the black hole, including simultaneous rotation in different planes, and the existence of \textit{ultra-spinning} regimes, which do not exist in $d=4$: recall that the unique stationary and axisymmetric black hole in $d=4$, the Kerr black hole \cite{KerrOriginal}, must satisfy $2a\leq\mu$ where $\mu$ is a mass parameter and $a$ is proportional to the ratio between the angular momenta and the mass. This upper bound  in the speed of rotation is referred to as the \textit{Kerr bound}. In this section we will see that such bound is not necessarily present in the generalisation to $d>4$ of the Kerr solution: the Myers-Perry (MP) black holes \cite{MPSolution}. The richness of such solutions is wide, and since its discovery several aspects have been studied, such as its global structure, hidden symmetries, stability, and more. In this section we will focus mainly on the novel features on the rotation dynamics that are not present in $d=4$.
\subsection{The general MP solution}
The MP solution describes spherical black holes rotating in, at most, $N=\lfloor\frac{d-1}{2}\rfloor$ planes. Although it might be regarded as the generalisation of the Kerr solution to $d>4$, the obtention of such solution is considerably more difficult than the ones presented so far. The solution with a single spin can be obtained in a reasonable simple manner motivating an appropriate ansatz via educated guesswork, but the case of multiple spins requires a deeper insight. The acute observation of Myers and Perry in \cite{MPSolution} is that the most natural way of extending the Kerr black hole to higher dimensions is to regard it as a \textit{Kerr-Schild spacetime}. In a Kerr-Schild spacetime local charts at every point exist such that
\begin{equation}\label{eq:42}
g_{\mu\nu}=\eta_{\mu\nu}+hk_{\mu}k_{\nu},
\end{equation}
being $h$ a function and $k^{\mu}:=g^{\mu\nu}k_{\nu}$ a vector which is null wrt the Minkowski metric $\eta_{\mu\nu}$ and, hence, also wrt $g_{\mu\nu}$. In other words, such class of spacetimes are given exactly by the "perturbations" around flat space. The Kerr black hole in $d=4$ belongs to this class of spacetimes, so an extension to $d>4$ in the form \eqref{eq:42} is indeed a natural choice. A remarkable property of such spaces is that, with an appropriate choice for the form of $k_{\nu}$, the Einstein equations become linear and consequently easier to integrate.

In the following we briefly sketch the procedure in \cite{MPSolution} towards the obtention of the general MP solution. The first step is to write the metric in the form \eqref{eq:42} for an undetermined function $h$, and propose a vector $k_{\mu}\text{d}x^{\mu}$ which has a similar form to that of Kerr when written in the Kerr-Schild gauge. Organise the spatial coordinates in pairs as $(x^{i},y^{i}):=(x^{i},x^{i+1})$ with the last coordinate unpaired in the even case, where $i$ runs from $1$ to $N$, and for even $d$, $z:=x^{d-1}$. Then, assuming summation (here and henceforth) over $i$ whenever it is repeated, the choice for $k_{\mu}$ is
\begin{align}\label{eq:43}
k_{\mu}\text{d}x^{\mu}=\text{d}t+\frac{r(x^{i}\text{d}x^{i}+y^{i}\text{d}y^{i})-a_{i}(x^{i}\text{d}y^{i}-y^{i}\text{d}x^{i})}{r^{2}+a_{i}^{2}}\,\,\,\,\,\,\,\,\text{(d odd)}\\ \nonumber
k_{\mu}\text{d}x^{\mu}=\text{d}t+\frac{r(x^{i}\text{d}x^{i}+y^{i}\text{d}y^{i})-a_{i}(x^{i}\text{d}y^{i}-y^{i}\text{d}x^{i})}{r^{2}+a_{i}^{2}}+\frac{z\text{d}z}{r} \,\,\,\,\,\,\,\,\text{(d even)}
\end{align}
where $a_{i}\in \mathbb{R}$ and $r(x^{i},y^{i})$ is a function that is fixed implicitly by the condition $0=\eta^{\mu\nu}k_{\mu}k_{\nu}$. After some manipulation, such condition reads
\begin{equation}\label{eq:44}
\frac{{x^{i}}^{2}+{y^{i}}^{2}}{r^{2}+a_{i}^{2}}=1\,\,\,\,\,\,\,\,\text{(d odd)},\,\,\,\,\,\,\,\,\,\,\,\,\,\,\,\,\,\,\frac{{x^{i}}^{2}+{y^{i}}^{2}}{r^{2}+a_{i}^{2}}+\frac{z^{2}}{r^{2}}=1 \,\,\,\,\,\,\,\,\text{(d even)}.
\end{equation}
We have reduced the problem to determining a single function $h$. Choosing an appropriate rigid, null tetrad (i.e. a tetrad such that the metric components are constant and at least one basis vector is null) one can compute the vacuum Einstein equations $R_{\alpha\beta}=0$ \footnote{Throughout this work, we use latin characters as abstract indices, late greek alphabet indices to label components in coordinate basis and early greek alphabet indices to label components in tetrad basis.} using the Cartan structure equations and solve them for $h$. We refer the reader to Appendix A in \cite{MPSolution} for the (very long) explicit calculations.
In terms of the analogue of the direction cosines introduced in the previous section, the general MP solution reads 
\begin{equation}\label{eq:45}
ds^{2}=-dt^{2}+(r^{2}+a_{i}^{2})(d\mu_{i}^{2}+\mu^{2}_{i}d\phi_{i}^{2})+\frac{\mu r^{2}}{\Pi F}(dt-a_{i}\mu_{i}^{2}d\phi_{i})^{2}+\frac{\Pi F}{\Pi-\mu r^{2}}dr^{2}
\end{equation}
for $d$ odd, and 
\begin{equation}\label{eq:46}
ds^{2}=-dt^{2}+r^{2}d\alpha^{2}+(r^{2}+a_{i}^{2})(d\mu_{i}^{2}+\mu^{2}_{i}d\phi_{i}^{2})+\frac{\mu r}{\Pi F}(dt-a_{i}\mu_{i}^{2}d\phi_{i})^{2}+\frac{\Pi F}{\Pi-\mu r}dr^{2} 
\end{equation}
for $d$ even. In both cases $\mu$ is a real parameter, 
\begin{equation}\label{eq:47}
F(r,\mu_{i})=1-\frac{a_{i}\mu_{i}^{2}}{r^{2}+a^{2}_{i}},\,\,\,\,\,\,\,\Pi=\prod_{i=1}^{N}(r^{2}+a_{i}^{2})
\end{equation}
and the coordinates satisfy the usual constraint $\mu_{i}^{2}=1$ or $\mu^{i}+\alpha^{2}=1$ for $d$ odd or even, respectively. For the asymptotic limit $r\to\infty$ the solution becomes, 
\begin{align}\label{eq:49}
ds^{2}=&-\left(1-\frac{\mu}{r^{d-3}}\right)dt^{2}-2\frac{\mu}{r^{d-3}}a_{i}\mu_{i}^{2}d\phi_{i}dt
&+\left(1+\mathcal{O}\left(\frac{1}{r^{d-3}}\right)\right)\left(dr^{2}+r^{2}(d\mu_{i}^{2}+\mu_{i}^{2}d\phi_{i}^{2})+r^{2}d\alpha^{2}\right)
\end{align}
with $\alpha=0$ in the case $d$ odd. Hence, the physical mass and angular momenta are given by
\begin{equation}\label{eq:49a}
M=\frac{(d-2)\Omega_{d-2}}{16\pi G}\mu,\,\,\,\,\,\, J=\frac{2}{d-2}Ma.
\end{equation}
Finally, it is interesting to mention that Kerr-Schild vacuum spacetimes (and, hence, MP black holes) are \textit{algebraically special spacetimes} for $d\geq4$ being precisely $k_{\mu}$ a multiple \textit{Weyl-aligned null direction} (WAND) which, in addition, turns out to be geodesic \cite{WandsKerrSchild}. However, unlike in the $d=4$ case, the congruence of integral curves of $k^{\mu}$ is \textit{not} shear-free. Unfortunately, in this work we will not deal with the algebraic classification of GR solutions in higher dimensions. 
\subsection{MP Black Holes with a single spin} 
 In order to capture the essence of the new qualitative behaviour of rotation in $d>4$, we will start with the case of a black whole with a single spin in an arbitrary number of dimensions. In order to put the solution in a familiar way we will use a new set of coordinates. It is convenient to define them first in flat space. Consider inertial coordinates $(t,x^{1},...,x^{d-1})$. Now organise them into two sets, a plain of rotation $(x^{1},x^{2})$, and the remaining ones $(x^{3},...,x^{d-1})$. Introduce polar coordinates $(r_{1},\phi)$ in the plane via $x^{1}=r_{1}\cos{\phi}$ and $x^{2}=r_{1}\sin{\phi}$, and spherical coordinates $(r_{2},\chi_{1},...,\chi_{d-4})$ (centred in certain spatial 2-surface) for the rest of the space. Now define an "axial" coordinate $\theta$ that controls the projection of $r$ into the plane of rotation and the direction specified by $\{\chi\}$\footnote{Notice that $\theta$ lives in $(0,\pi/2)$ rather than $(0,\pi)$ which indicates it is not strictly speaking an axial angle. The reason for this is that now $r_{2}=r\cos\theta$ must remain positive.},
\begin{equation}\label{eq:9}
r_{1}=r\sin\theta,\,\,\,\,\,\,\,\,\,\,r_{2}=r\cos\theta.
\end{equation}
Eliminating $r_{1}$ and $r_{2}$ with $r$ and $\theta$ flat space reads
\begin{equation}\label{eq:10}
ds^{2}=-dt^{2}+dr^{2}+r^{2}(d\theta^{2}+\sin^{2}\theta d\phi^{2})+r^{2}\cos^{2}\theta d\Omega_{d-4}^{2}.
\end{equation}
Under the appropriate coordinate transformation, \eqref{eq:45} and \eqref{eq:46} read
\begin{align}\label{eq:20}
ds^{2}=&-dt^{2}+\frac{\mu }{r^{d-5}\Sigma}(dt-a\sin^{2}\theta d\varphi)^{2}+\frac{\Sigma}{\Delta}dr^{2}+\Sigma d\theta^{2}+(r^{2}+a^{2})\sin^{2}\theta d\varphi^{2}+r^{2}\cos^{2}\theta d\Omega^{2}_{d-4}
\end{align}
with
\begin{equation}\label{eq:19}
\Delta(r)=r^{2}+a^{2}-\frac{\mu}{r^{d-5}},\,\,\,\,\,\,\,\, \Sigma=r^{2}+a^{2}\cos^{2}\theta.
\end{equation}
For vanishing $\mu$ and $a$, \eqref{eq:19} becomes \eqref{eq:10}. Consequently, surfaces of constant $r$ are topological $(d-2)$-spheres.
\subsubsection{Ultra-spinning Black Holes}
For the moment, the MP black hole \eqref{eq:20} looks very similar to the Kerr black hole. The new remarkable properties arise from the explicit dependence on $d$. As one would expect from comparison to its four dimensional counterpart, \eqref{eq:20} exhibits an event horizon at the largest real root (if any), $r_{H}$, of the equation $\Delta(r_{H})=0$ \cite{MPSolution}. Then, the topology of the horizon is automatically $S^{d-2}$. Let us now study when \eqref{eq:20} presents a regular horizon. Explicitly, $\Delta(r_{H})=0$ reads
\begin{equation}\label{eq:23}
r_{H}^{2}+a^{2}-\frac{\mu}{r_{H}^{d-5}}=0.
\end{equation}
First of all, notice that such equation has no solution in $r>0$ if $\mu<0$. From the study in \cite{MPSolution}, curvature singularities occur at $r\leq0$. Then, in order to avoid naked singularities we will restrict ourselves to $\mu>0$. For $d=5$, \eqref{eq:23} is solved exactly by $r_{H}=\sqrt{\mu-a^{2}}$ and we are again in front of a Kerr bound: if $a>\sqrt{\mu}$ the solution in nakedly singular. If $a=\sqrt{\mu}$ we will see that the "transverse $(d-4)$-spheres" of the horizon shrink and get zero area leading again to naked singularities. For $d\geq6$, \eqref{eq:23} always has a solution in $r>0$ for all $\mu>0$ and for all $a$, due to the divergence as $r\to0$ of dimensional dependent term $\frac{-\mu}{r^{d-5}}$. In other words, there is no upper bound in the angular momentum for any $\mu>0$. Conventionally, black hole solutions living in the region of the parameter space in which $a\gg\mu$ are called \textit{ultra-spinning}.

\begin{figure}[h!]
\centering
\includegraphics[scale=0.4]{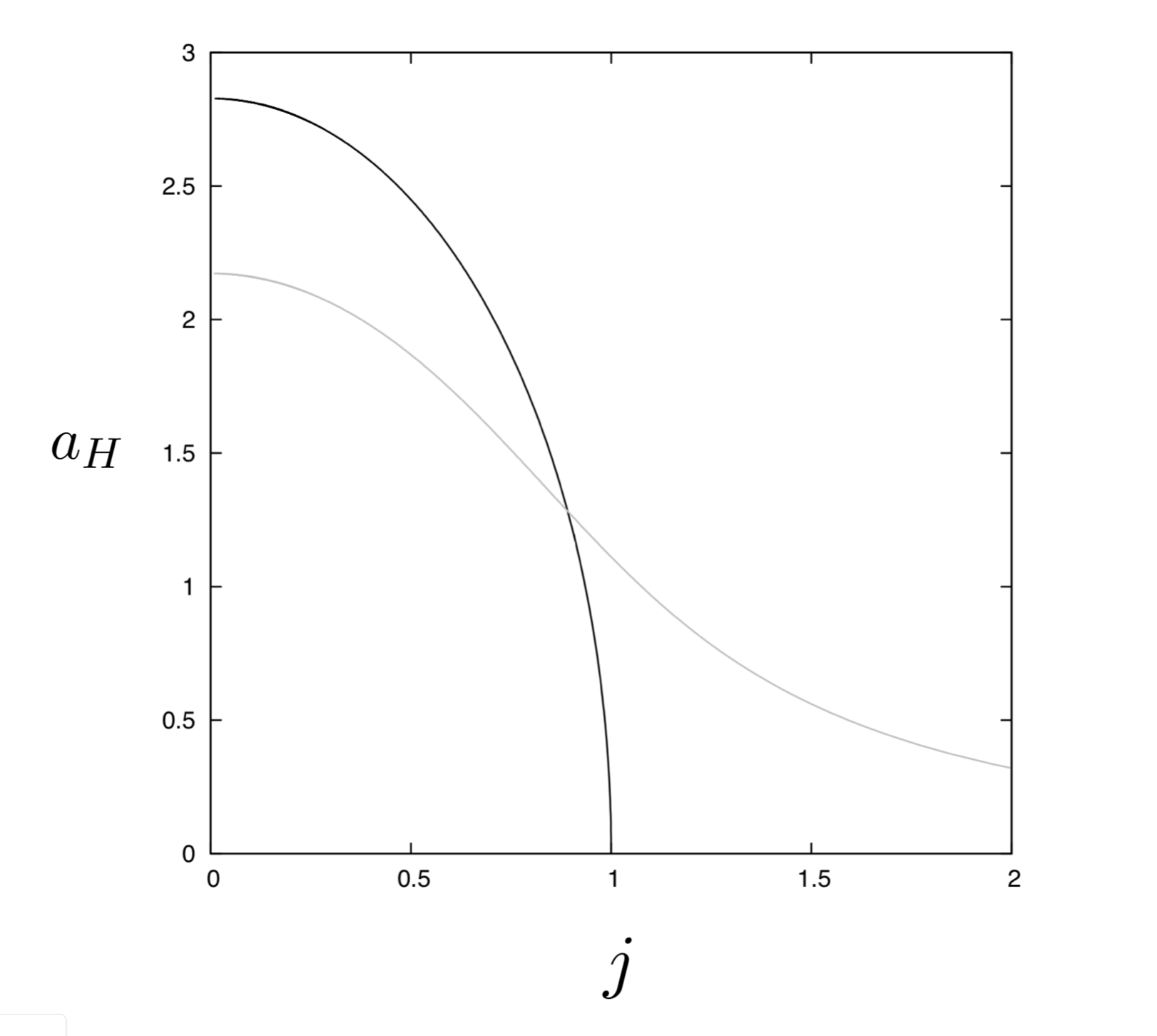}
\caption{Curve $a_{H}(j)$ for $d=5$ (black) $d=6$ (gray).}
\label{fig6}
\end{figure}

For many reasons that will become clear throughout this work, it is interesting to study the behaviour of the horizon area in the different rotation configurations. For the moment we just have a single spin so this reduces to obtaining the curve $a_{H}(j)$. After some straightforward manipulation, the pull back of the metric on (the spatial cross sections of) the horizon reads
\begin{equation}\label{eq:25}
ds^{2}_{H}=\frac{\left(r_{H}^{2}+a^{2}\right)^{2}}{r_{H}^{2}+a^{2}\cos^{2}\theta}\sin^{2}\theta d\phi^{2}+(r_{H}^{2}+a^{2}\cos^{2}\theta)d\theta^{2}+r^{2}_{H}\cos^{2}\theta d\Omega^{2}_{d-4}
\end{equation}
where we have used \eqref{eq:23}. The area of the horizon can now be computed as\footnote{Recall from the discussion about \eqref{eq:10} that $\theta$ is not a standard axial coordinate, and its range goes from $0$ to $\pi/2$ only.}
\begin{align}\label{eq:26}
\mathcal{A}_{H}=&\int_{0}^{2\pi}\text{d}\phi\int_{0}^{\pi/2}\text{d}\theta\int\text{d}\Omega_{d-4}r_{H}^{d-4}\cos^{d-4}\theta\left(r^{2}_{H}+a^{2}\right)\sin\theta=\frac{2\pi\left(r^{2}_{H}+a^{2}\right)}{d-3}r^{d-4}_{H}\Omega_{d-4}
\end{align}
where $\text{d}\Omega_{d-4}$ is the volume element on the unit round $(d-4)$-sphere, and the last integral indicates integration along the full space of $d-4$ angles $\{\chi\}$. This is a particularly convenient way of writing the area for our following discussion, but an alternative form is obtained using the recurrence relation $\Omega_{d-1}=\frac{d}{2\pi}\Omega_{d+1}$,
\begin{equation}\label{eq:27}
\mathcal{A}_{H}=r_{H}^{d-4}(r_{H}^{2}+a^{2})\Omega_{d-2}.
\end{equation} 
In order to obtain $a_{H}(j)$, one could try to write $\mathcal{A}_{H}$ as an explicit function of $a$, and then just use \eqref{eq:49a} and the definition of $j$. In our case this is not possible as we can not solve explicitly \eqref{eq:23} for $r_{H}(a)$. However, we can solve if for $a(\nu)$, being $\nu:=r_{H}/a$ a dimensionless parameter,
\begin{equation}\label{eq:4646}
a=\mu^{\frac{1}{d-3}}\left(\frac{\nu^{5-d}}{\nu^{2}+1}\right)^{\frac{1}{d-3}}.
\end{equation}
Hence, the area can be written explicitly as a function of $\nu$, $\mathcal{A}_{H}(\nu)$. Now it is just a matter of going through the definitions to get $a_{H}(\nu)$. Following a similar procedure one obtains $j(\nu)$, and $a_{H}(j)$ can be given in parametric form,
\begin{align}\label{eq:29}
a_{H}^{d-3}=&8\pi\left(\frac{d-4}{d-3}\right)^{\frac{d-3}{2}}\frac{\Omega_{d-3}}{\Omega_{d-2}}\frac{\nu^{2}}{1+\nu^{2}},\,\,\,\,\,\,\,\,\,
j^{d-3}=\frac{\pi}{(d-3)^{\frac{d-3}{2}}}\frac{\Omega_{d-3}}{\Omega_{d-2}}\frac{\nu^{5-d}}{1+\nu^{2}}.
\end{align}
It is important to identify the different regimes of rotation along the parameter space $0<\nu<\infty$. Looking at $j(\nu)$, we observe that for $d\geq6$ the ultra-spinning solutions are located at $\nu\to0$, while the static solutions with $j=0$ are recovered as $\nu\to\infty$. The area at $\nu\to\infty$ becomes that of the Schwarzschild-Tangherlini black hole, while in the ultra-spinning regime $\nu\to0$ it decreases as $\sim\nu^{\frac{2}{d-3}}$. That is, increasing the spin an \textit{infinite} amount would produce the net effect of 'shrinking' the black hole to zero area. For $d=5$ the static regime is also recovered as $\nu\to\infty$, and in such case the area approaches the spherically symmetric value. However, regarding the fast rotation regime, we see that $j(\nu)$ is monotonically decreasing with a global maximum at $\nu=0$, as a consequence of the Kerr bound. In such case, the area vanishes, $a_{H}(0)=0$, so the horizon is pathological when the Kerr bound gets saturated. The significant difference with the case $d\geq6$ is that now it takes only a \textit{finite} amount of spin to make the black hole shrink to null area. The content of this discussion is encoded in Figure \eqref{fig6}.
\subsubsection{Horizon geometry in the ultra-spinning regime}
There are other more subtile features of the MP black holes that are not present in four dimensions. In this section we are interested in two of them. First, in $d=4$ the horizon area of the Kerr black hole decreases for increasing rotation, but it is always nonzero and larger than the area of the maximally rotating case, $\mathcal{A}_{H}\sim\mu^{2}$. The second feature is that the Kerr black hole is always confined within a finite radius in the rotation plane. Let us give a more precise notion of this. Consider a single-spin MP black hole with horizon geometry given by \eqref{eq:25}. We define the \textit{confinement radius} as
\begin{equation}\label{000}
R^{(d)}_{\pi/2}:=\frac{\mathfrak{l}^{(d)}_{\pi/2}}{2\pi},
\end{equation}
where $\mathfrak{l}^{(d)}_{\pi/2}$ is the proper length of the equator line of the horizon in the plane of rotation\footnote{That is, the length of a closed loop tangent to $\partial_{\phi}$ with $\theta=\pi/2$ and $\chi$ constant.}, and $d$ is the spacetime dimension. For the Kerr black hole we have that $R_{\pi/2}^{(4)}\leq\mu$, which gets saturated at the Kerr bound. That is, the black hole is always confined within $R_{\pi/2}^{(4)}=\mu$, and the 'centrifugal force' can not 'pancake it' more than that.

In $d>4$, both properties mentioned above are violated in different ways. Let us start with the area. In $d=5$, $\mathcal{A}_{H}$ strictly vanishes when the spin reaches the Kerr bound, and in $d\geq6$ it can get arbitrarily close to zero for arbitrarily large spin. A first hint towards understanding this new behaviour can be extracted writing the horizon area in the form
\begin{equation}\label{eq:30}
\mathcal{A}_{H}=\left(\frac{2\pi\left(r^{2}_{H}+a^{2}\right)}{d-3}\right)\left(r^{d-4}_{H}\Omega_{d-4}\right).
\end{equation}
We observe that $\mathcal{A}_{H}$ has two contributions: the first factor in \eqref{eq:30} is a Kerr-looking one, while the second one is exactly the area of a $(d-4)-$dimensional round sphere of radius $r_{H}$. In the fast rotation regimes (that is, $a\to\sqrt{\mu}$ for $d=5$ and $a\to\infty$ for $d\geq6$) we learn from \eqref{eq:23} that $r_{H}$ becomes
\begin{equation}\label{eq:31}
r_{H}=\begin{cases}\sqrt{\mu-a^{2}} &d=5\\ \sim\left(\frac{\mu}{a^{2}}\right)^{\frac{1}{d-5}}&d\geq6\end{cases},
\end{equation}
so $r_{H}$ vanishes identically at the Kerr bound in $d=5$, and converges to zero for $d\geq6$ as $a\to\infty$. Hence, it is the second 'spherical' factor in \eqref{eq:30} that makes $\mathcal{A}_{H}$ vanish and decrease arbitrarily in $d=5$ and $d\geq6$, in the respective fast rotation regimes. This suggests that, while the area of the Kerr factor "parallel" to the rotation plane remains finite, the area of the "transverse" $(d-4)$-dimensional spheres of the horizon goes to zero. This point of view about the horizon area was first studied in \cite{Ultraspinning} for $d\geq6$. Here we follow their discussion, but also including $d=5$ and studying the behaviour of the confinement radius \eqref{000}. First of all, notice that $r_{H}$ has no physical meaning as it is not intrinsically-defined, so we have to reformulate the notions of transverse and parallel areas in an invariant form. We define the \textit{parallel area} $\mathcal{A}_{\parallel}$ as the area of a section $\chi=\chi_{0}$ of the horizon, where $\chi_{0}$ are $d-4$ cosntants. Using \eqref{eq:25} we have
\begin{equation}\label{eq:32}
\mathcal{A}_{\parallel}=\int_{0}^{2\pi}\text{d}\phi\int_{0}^{\pi/2}\text{d}\theta(r_{H}^{2}+a^{2})\sin\theta=\frac{1}{2}(r_{H}^{2}+a^{2})\Omega_{2}.
\end{equation}
Notice that in $d=5$ the parallel area is
\begin{equation}\label{eq:A6}
\mathcal{A}_{\parallel}=\frac{1}{2}\mu\Omega_{2}
\end{equation}
so it is independent of the spin $a$. Now define the \textit{transverse area} $\mathcal{A_{\perp}}$ as the area of a section $\theta=\theta_{0}$ and $\phi=\phi_{0}$ of the horizon, being $\theta_{0}$ and $\phi_{0}$ constants. Again using \eqref{eq:25},
\begin{equation}\label{eq:33}
\mathcal{A}_{\perp}=\int \text{d}\Omega_{d-4} r_{H}^{d-4}\cos^{d-4}\theta_{0}=\left(r_{H}\cos\theta_{0}\right)^{d-4}\Omega_{d-4}.
\end{equation}
Let us now study the behaviour of these magnitudes in the fast rotation regime. From \eqref{eq:31} we see that in such regime $\nu=\frac{r_{H}}{a}\ll1$ for both $d=5$ and $d\geq6$. Then, the different areas defined above become
\begin{align}\label{eq:35}
\mathcal{A}_{\parallel}\sim a^{2},\,\,\,\,\,\,\,\,\,\,\, \mathcal{A}_{\perp}\sim r_{H}^{d-4}, \,\,\,\,\,\,\,\,\,\,\, \mathcal{A}_{H}\sim a^{2}r_{H}^{d-4}.
\end{align}
This confirms our prediction: the area of the transverse spheres $\mathcal{A}_{\perp}$ vanishes identically in the Kerr bound for $d=5$, and can become arbitrarily close to zero for $d\geq6$ in the ultra-spinning regime. Regarding $\mathcal{A}_{\parallel}$, we see that in $d\geq6$ it can become arbitrarily large as $a\to\infty$. As pointed out above, in $d=5$ the parallel area is given by \eqref{eq:A6} for all $a$, so it remains finite also in the Kerr bound. In addition, notice that the first two equations in \eqref{eq:35} provide a length scale for the parallel area $\ell_{\parallel}=a$ and the transverse area $\ell_{\perp}=r_{H}$ respectively, and they give the expected scale for the total area $\mathcal{A}_{H}\sim \ell_{\parallel}^{2}\ell_{\perp}^{d-4}$.

In $d\geq6$ we can give an heuristic interpretation of these results: in contrast to what happens in $d=4$, now the centrifugal force can pancake the black hole infinitely along the plane of rotation. For $d=5$, however, the fact that $\mathcal{A}_{\parallel}$ remains finite suggests that the black hole can not be infinitely spread along the plane of rotation. Rather surprisingly, that is not the case. To see this, compute the confinement radius for any $d$,
\begin{equation}\label{eq:36}
R^{(d)}_{\pi/2}=a\frac{(\nu^{2}+1)}{\nu},
\end{equation} 
where recall $\nu=r_{H}/a$. Although \eqref{eq:36} does not show explicit dependence in $d$, the range in which $\nu$ lives is different for $d=4$ and $d\geq5$: for $d\geq5$, we have $0<\nu<\infty$ as can be deduced, for example, from \eqref{eq:31}. However, in $d=4$ 
\begin{equation}\label{eq:37}
\nu=\frac{\mu}{2a}+\sqrt{\left(\frac{\mu}{2a}\right)^{2}-1}
\end{equation}
so in this case $1<\nu<\infty$. Studying $R^{(d)}_{\pi/2}$ in the fast spinning regimes ($\nu\to0$ and $\nu\to1$ for $d\geq5$ and $d=4$ respectively), we get
\begin{equation}\label{eq:38}
R^{(d)}_{\pi/2}\to\begin{cases}\mu & d=4 \\ \infty & d\geq5\end{cases}.
\end{equation}
Nothing is surprising for $d=4$ and $d\geq6$, as the parallel areas are finite and infinite respectively in the fast rotation regime. For $d=5$, however, although the parallel area remains always finite, close to the Kerr bound the black hole is spread infinitely along the plane of rotation, according to \eqref{eq:38}. In other words, it approaches the shape of an extended black object like in the case $d\geq6$, but now the spin only has to be increased by a finite amount. 
\begin{figure}[h!]
\centering
\includegraphics[scale=0.5]{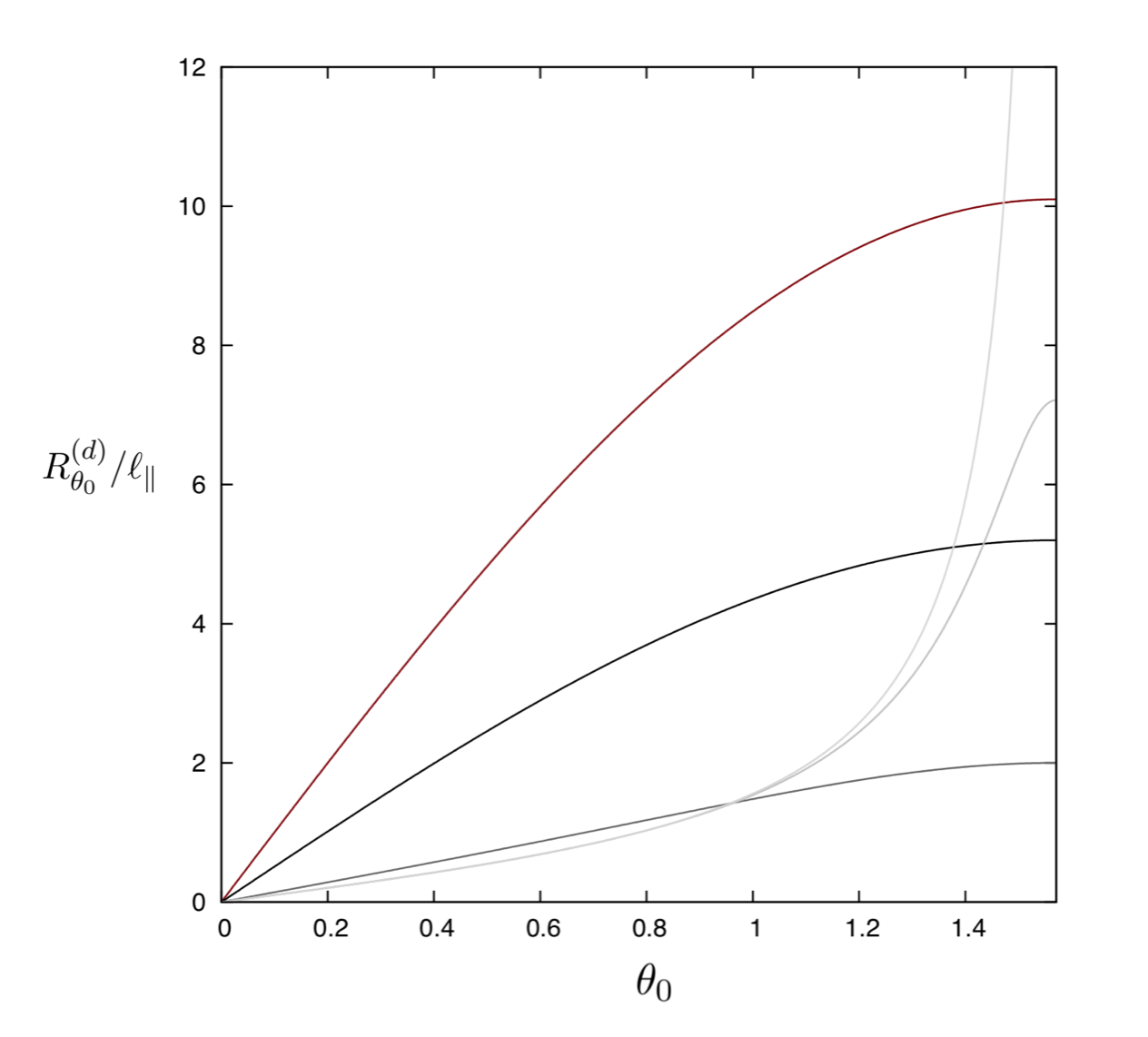}
\caption{Relation $R^{(d)}_{\theta_{0}}/\ell_{\parallel}(\theta_{0})$ for spherical symmetry (maroon), and subsequently decreasing values of $\nu$ for decreasing darkness of the lines. The Kerr bound in $d=4$ corresponds to the flattest line, with $\nu=1$.}
\label{fig0}
\end{figure}
Finally, it is worth nothing to generalise the definition of the confinement radius now allowing the 'axial' angle $\theta_{0}$ to take different values other than $\pi/2$: denoting $\mathfrak{l}^{(d)}_{\theta_{0}}$ the proper length of a closed loop with tangent $\partial_{\phi}$ and $\theta=\theta_{0}$, $\chi=\chi_{0}$ we define $R^{(d)}_{\theta_{0}}:=\mathfrak{l}^{(d)}_{\theta_{0}}/2\pi$. Using \eqref{eq:25} it is immediate to get
\begin{equation}\label{eq:39}
R^{(d)}_{\theta_{0}}=a(1+{\nu}^{2})\frac{\sin\theta_{0}}{\sqrt{\nu^{2}+\cos^{2}\theta_{0}}}.
\end{equation}
Plotting $R^{(d)}_{\theta_{0}}/\ell_{\parallel}$ as a function of $\theta_{0}$  for different values of $\nu$ gives a visual picture of how the shape of the black hole varies in different regimes of rotation. The results are shown in Figure \eqref{fig0}.

\begin{figure}[h!]
\centering
\includegraphics[scale=0.4]{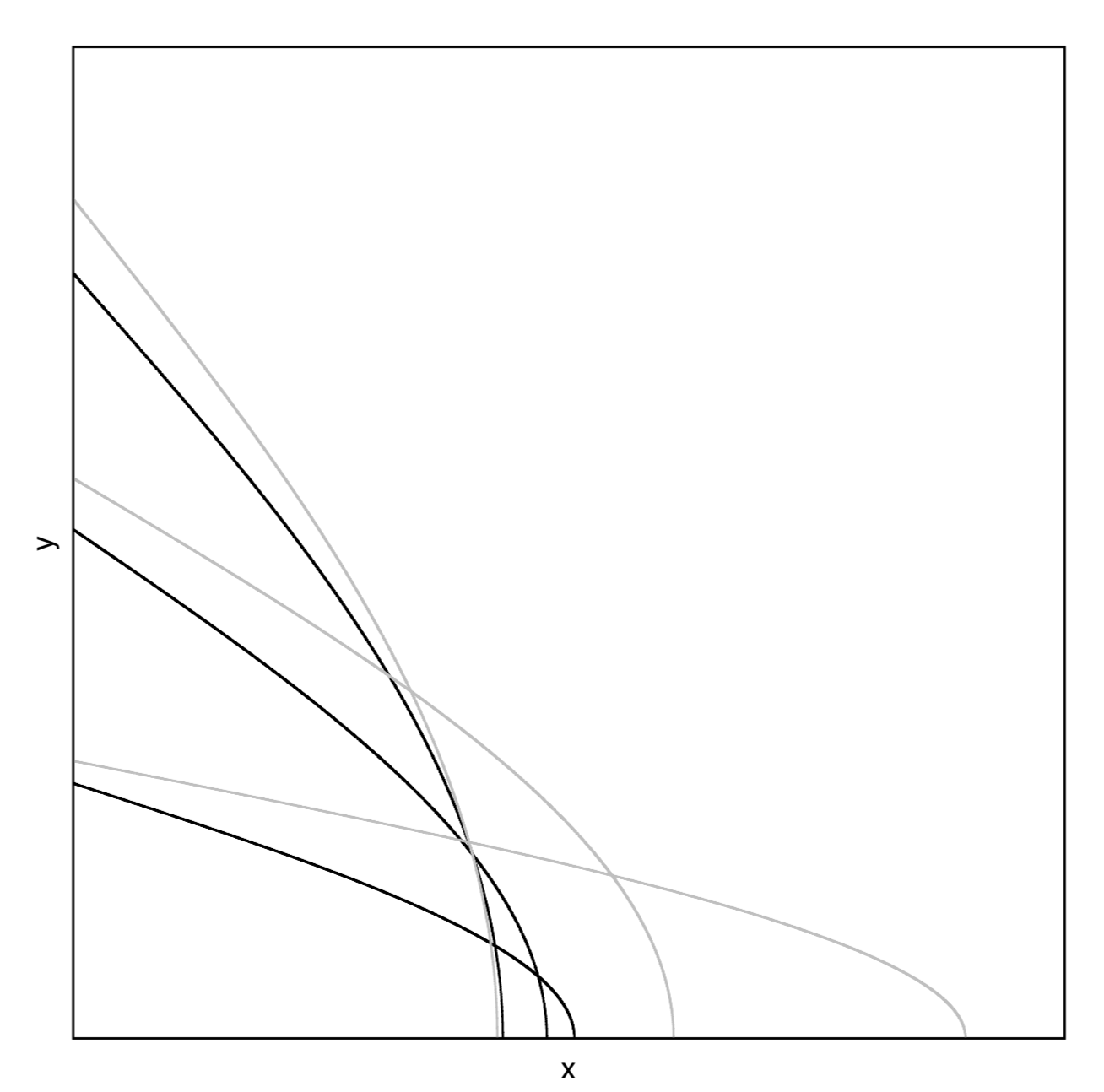}
\caption{Isometric embedding \eqref{eq:A16} in $d=5$ (black) and $d=6$ (grey). In both cases, we represent curves belonging to horizons with areas $a_{H}=1.5,1.0,0.5$. The corresponding values of $\nu$ are $\nu=0.63, 0.38, 0.18$ in $d=5$, and $\nu=0.7,0.33,0.11$ in $d=6$. The flatter is the curve, the lower are $\nu$ and $a_{H}$.}
\label{figA}
\end{figure}

A nicer way of visualising how the black hole's shape behaves is via an \textit{isometric embedding}\footnote{The part of this section having to do with isometric embeddings has been added to this essay for completeness after handing it in for evaluation during Part III. It was not part of the original version.}. Assume we have a 2-dimensional riemannian manifold $H$ with metric tensor field $g$. Consider the 3-dimensional euclidean space $\mathbb{E}^{3}$, with euclidean metric $\delta$. Then, $H$ can be isometrically embedded in $\mathbb{E}^{3}$ if there exists an embedding $\Phi:H\longrightarrow \mathbb{E}^{3}$ such that
\begin{equation}\label{eq:A8}
\Phi^{*}\delta=g.
\end{equation}
We want to embed the sections of constant $\chi$ of the event horizon \eqref{eq:25} in $\mathbb{E}^{3}$. In coordinates $(\theta,\phi)$, the geometry of such sections reads
\begin{equation}\label{eq:A9}
ds^{2}_{H}=\frac{\left(r_{H}^{2}+a^{2}\right)^{2}}{r_{H}^{2}+a^{2}\cos^{2}\theta}\sin^{2}\theta d\phi^{2}+(r_{H}^{2}+a^{2}\cos^{2}\theta)d\theta^{2}.
\end{equation}
Since $g_{\mu\nu}$ in \eqref{eq:A9} do not depend on $\phi$, we can try with an embedding of the form
\begin{equation}\label{eq:A11}
x(\theta,\phi)=f(\theta)\cos{\phi},\,\,\,\,\, y(\theta,\phi)=f(\theta)\sin{\phi},\,\,\,\,\, z=h(\theta)
\end{equation}
for some functions $f(\xi)$ and $h(\xi)$, where $(x,y,z)$ are the standard euclidean coordinates. Then, equation \eqref{eq:A8} leads to
\begin{equation}\label{eq:A12}
f(\theta)=\sqrt{g_{\phi\phi}},\,\,\,\,\, \left(h'(\theta)\right)^{2}=g_{\theta\theta}-\left(\partial_{\theta}\sqrt{g_{\phi\phi}}\right)^{2}.
\end{equation}
Not all the riemannian surfaces can be embedded isometrically in $\mathbb{E}^{3}$. For instance, from the second equation in \eqref{eq:A12} we obtain a necessary condition for the existence of the isometric embedding of our surface, which is
\begin{equation}\label{eq:A13}
1\leq\frac{g_{\theta\theta}}{\left(\partial_{\theta}\sqrt{g_{\phi\phi}}\right)^{2}}=\frac{1}{\cos^{2}\theta}\left(\frac{\nu^{2}+\cos^{2}\theta}{\nu^{2}+1}\right)^{4}.
\end{equation}
Unfortunately, with standard arguments one can easily check that such condition is not satisfied for all $\nu$ and $\theta$. For instance, the ultra spinning regime in $d\geq5$ lies in $0<\nu<1$ and in such range of parameters the right hand side of \eqref{eq:A13} is smaller than one for an interval of $\theta$ within $(0,\pi/2)$, so the surface can not be fully embedded. However, for slowly rotating black holes with $\nu\gg1$ the right hand side becomes $1/\cos^{2}\theta$ so the surface can be fully embedded in that case. Then, one can integrate \eqref{eq:A12} in such limit and observe the small deformation of the black hole around staticity as its angular momenta increases. Here, however, we are interested in the ultra spinning regimes and we have seen that this embedding does not work in such case. Nevertheless, in an act of stubbornness one can still find ways of representing isometrically the deformation of the black hole in the ultra spinning regime, and actually in all the range of $\nu$. Instead of the surface \eqref{eq:A9}, consider just the curves $\phi=\phi_{0}$, with $0<\theta<\pi/2$. Their geometry reads
\begin{equation}\label{eq:A14}
ds^{2}_{H,\phi_{0}}=\left(r_{H}^{2}+a^{2}\cos^{2}\theta\right)d\theta^{2},
\end{equation} 
and now the idea is to isometrically embed this curve in $\mathbb{E}^{2}$ as described above. Using standard euclidean coordinates $(x,y)$ in $\mathbb{E}^{2}$, the analogue of equation \eqref{eq:A8} now leads to
\begin{equation}\label{eq:A15}
\left(\frac{dx}{d\theta}\right)^{2}+\left(\frac{dy}{d\theta}\right)^{2}=r_{H}^{2}+a^{2}\cos^{2}\theta,
\end{equation}
which is solved by
\begin{equation}\label{eq:A16}
x=a\sin\theta,\,\,\,\,\,\,\,\,\, y=r_{H}\left(\frac{\pi}{2}-\theta\right).
\end{equation}
A plot of the isometric embedding \eqref{eq:A16} for different values of $\nu$ in $d=5$ and $d=6$ is given in Figure \eqref{figA}. We see in such curves how the black hole is spread along the plane of rotation as its spin increases, i.e. as $\nu\to0$. 

Here a word must be said about this result. Although this is an isometric embedding and, hence, it reproduces exactly the geometry of the curves of the horizon, the plots in Figure \eqref{figA} can \textit{not} be regarded as isometric cross sections of the surface \eqref{eq:A9}: for instance, when considering the static limit $a\to0$ the embedded curve \eqref{eq:A16} becomes a straight line instead of a quarter of a circle, and Figure \eqref{figA} exhibits a peak in the north pole of the horizon but we know that the axis of rotation of the black hole is smooth and has no conical singularities. This is a consequence of the fact that, in general, there is no unique way of embedding a curve isometrically in euclidean space\footnote{For example, a curve of constant $\phi$ and $0<\theta<\pi$ of a unit round 2-sphere can be embedded in $\mathbb{E}^{2}$ as a straight line but also as a semicircle.}. Obtaining an embedding corresponding to an isometric cross section of the surface implies solving the problem of embedding the whole surface, and above we have seen that in our case such task is pathological.
\subsubsection{The membrane limit and the ultra-spinning Kerr bound}
Consider a single spin MP black hole in $d\geq6$. We have seen that for large $a$, the black hole is spread to infinity along the rotation plane. Therefore, it is interesting to see up to what point an ultra-spinning black hole behaves like an extended black object, such as a black membrane. Here we follow \cite{Ultraspinning} to define a limit geometry for \eqref{eq:20} in the \textit{infinitely fast} rotation regime. We want such limit geometry to contain a black hole, so the limit has to be taken in such a way that an horizon remains present. Looking at \eqref{eq:31}, we see that $r_{H}$ is kept finite when $a\to \infty$ if $\mu\to\infty$ in such a way that $i)$ $\hat{\mu}=\mu/a^{2}$ remains finite. If taking into account $i)$ we look at $g_{t\varphi}$ in \eqref{eq:20}, we see that $g_{t\varphi}=-a\frac{\hat{\mu}}{r^{d-5}\cos^{2}\theta}\sin^{2}\theta$. This becomes infinitely large for $a\to\infty$ unless we focus on the central region of the horizon around $\theta=0$. Hence, the second condition for our limit is that $\theta\to0$ as $a\to\infty$ in such a way that $ii)$ the new coordinate $\sigma:=a\sin\theta$ is kept finite. Conditions $i)$ and $ii)$ fully determine our limit. Before taking it, it is instructive to perform an expansion of \eqref{eq:20} in powers of the infinitesimal factors $\frac{r}{a}$ and $\frac{\sigma}{a}$,
\begin{align}\label{eq:40}
g_{tt}&=-\left(1+\frac{\hat{\mu}}{r^{d-5}}\right)+\mathcal{O}\left(\frac{r^{2}}{a^{2}}\right)+\mathcal{O}\left(\frac{\sigma^{2}}{a^{2}}\right) \\ \nonumber
g_{rr}&=\frac{1}{1-\frac{\hat{\mu}}{r^{d-5}}}+\mathcal{O}\left(\frac{r^{2}}{a^{2}}\right)+\mathcal{O}\left(\frac{\sigma^{2}}{a^{2}}\right)\\ \nonumber
g_{t\varphi}&=\frac{-\hat{\mu}\sigma}{r^{d-5}}\frac{\sigma}{a}+\mathcal{O}\left(\frac{\sigma^{2}}{a^{2}}\right)=\mathcal{O}\left(\frac{\sigma}{a}\right)\\ \nonumber
g_{\varphi\varphi}&=\sigma^{2}+\mathcal{O}\left(\frac{r^{2}}{a^{2}}\right)+\mathcal{O}\left(\frac{\sigma^{2}}{a^{2}}\right)\\ \nonumber
g_{\sigma\sigma}&=\frac{g_{\theta\theta}}{a^{2}\cos^{2}\theta}=1+\mathcal{O}\left(\frac{r^{2}}{a^{2}}\right)+\mathcal{O}\left(\frac{r^{2}}{a^{2}}\frac{\sigma^{2}}{a^{2}}\right)
\end{align}
In our limit the terms of which we only specify the order in $r/a$ and $\sigma/a$ are identically zero so the geometry becomes
\begin{equation}\label{eq:41}
ds_{lim}^{2}=-\left(1-\frac{\hat{\mu}}{r^{d-5}}\right)dt^{2}+\frac{dr^{2}}{1-\frac{\hat{\mu}}{r^{d-5}}}+r^{2}d\Omega_{d-4}^{2}+d\sigma^{2}+\sigma^{2}d\varphi^{2},
\end{equation}
which is the black membrane\footnote{'$\text{Sch}_{d}$' means Schwarzschild-Tangherlini in $d$ dimensions.} $\text{Sch}_{d-2}\times\mathbb{R}^{2}$. Notice that, in this limit, rotation is not perceivable. Heuristically, the existence of the limit \eqref{eq:41} suggests that the infinitely fast rotating MP black holes are pancaked along the plane in which the $S^{d-2}$ horizon rotates, until it breaks in that direction adopting the black hole topology $S^{d-4}\times\mathbb{R}^{2}$.

Finally, although \eqref{eq:41} has been obtained as a limit of infinite rotation, the detailed expansion in \eqref{eq:40} reveals that it gives a good approximation of the geometry in the region around $\theta=0$ and close to the horizon of fast, but finitely rotating black holes. Indeed, take $a$ very large but finite. Then, from \eqref{eq:31} we know that $r_{H}/a\ll1$, so in the region in which $r$ is close to $r_{H}$ the terms $\mathcal{O}(r^{2}/a^{2})$ in \eqref{eq:40} remain small. If, in addition, we focus on the region around $\theta=0$, so that $\sigma\ll a$, then also the terms $\mathcal{O}(\sigma/a)$ and $\mathcal{O}(\sigma^{2}/a^{2})$ remain small. Therefore, close to the horizon in the region around the rotation axis of a finitely ultra-spinning black hole, the membrane \eqref{eq:41} approximates accurately the spacetime geometry. We have seen that such ultra-spinning black holes become largely spread along the plane of rotation, while the transverse spheres shrink, and furthermore the near horizon geometry resembles that of a black string. This suggests that ultra-spinning black holes might be unstable against GL-like perturbations. One can regard such possible instability as a natural \textit{Kerr bound for ultra-spinning black holes}.
\subsection{MP Black Holes with multiple spins}
In this section, we study the rotation dynamics of the general MP black hole solutions. Now, in addition to the ultra-spinning regimes discovered above, we will have the possibility of considering different spin configurations. 
\subsubsection{Ultra-spinning regimes and phase spaces}
The MP solutions are fully determined by $N+1$ parameters. For a given mass which, in our conventions, sets the scale of the rest of magnitudes, the parameter space in which the solution lives is $(j_{1},...,j_{N})$. The aim of this section is to study in which region of such space the MP solutions are physical (in the sense they do not have naked singularities) and exhibit a regular event horizon. That is, we want to obtain the \textit{phase space} of the solution.

The first step is to determine where is the horizon located. The metrics \eqref{eq:45} and \eqref{eq:46} are singular where 
\begin{equation}\label{eq:50}
i)\,\,\, \Pi F r^{-\gamma}=0,\,\,\,\,\,\,\,\,\,\,\,\,\, ii)\,\,\, \Pi-\mu r^{\gamma}=0,
\end{equation}
being $\gamma=2$ or $1$ for $d$ odd or even, respectively. Studying the analytical prolongation of \eqref{eq:45} and \eqref{eq:46}, one can see that $i)$ corresponds to physical curvature singularities, while $ii)$ is satisfied along an event horizon \cite{MPSolution,HorowitzBook}. Therefore, in order to construct the phase diagram we are mainly concerned with the solutions to the latter equation. For its study, it is convenient to define the polinomial 
\begin{equation}\label{eq:51}
\mathcal{P^{(\gamma)}}(r):=\Pi-\mu r^{\gamma},
\end{equation}
which has degree $2N$ (see \eqref{eq:47}), and $\gamma$ works as before. Solving $ii)$ in \eqref{eq:50} explicitly is not possible in general. Hence, in both cases $\gamma=1$ and $2$ we will deduce first some general results, and then specify spin configurations in which the presence of a regular horizon is guaranteed.

Let us first consider the case $d$ even. The first general property is that the mass parameter must satisfy $\mu>0$: since singularities appear at $r=0$, we need a solution in $r>0$ in order to avoid naked singularities. Given that the polynomial $\Pi$ and its derivative are everywhere non negative and monotonically increasing in $r$, it will only intersect the straight line $\mu r$ if $\mu>0$. Notice this is only a necessary condition for having a physical solution. Secondly, we have that there are $0$, $1$ or $2$ horizons: this can be seen first by checking that $\mathcal{P^{(\gamma)}}$ has exactly one minimum for both $\gamma=1,2$. Furthermore, $\mathcal{P^{(\gamma)}}$ grows as $r^{2N}$ for large $r$. A curve of this characteristics can only intersect a line of zero slope (and, in particular, vanish) at 0, 1 or 2 points. Let us now consider the particular configurations in which a regular horizon is guaranteed. Consider the spin configuration in which $m$ spins vanish with $m\geq1$. Then there is always a regular horizon regardless of the value of the rest of the spins: this is immediately seen writing $\mathcal{P}^{(1)}(r)$ for small $r$ in this configuration\footnote{We can assume wlog that the spins that vanish are the $m$ first ones.},
\begin{equation}\label{eq:53}
\mathcal{P}^{(1)}(r)\approx r^{2m}\left(\prod_{i=m+1}^{N}a_{i}^{2}\right)-\mu r
\end{equation} 
which is negative for small enough $r$ if $m\geq1$. Since at the same time $\mathcal{P}^{(1)}$ is positive and large for large $r$, it must vanish necessarily at some $r>0$. In particular, this implies that if one or more of the spins vanish, then the rest can be arbitrarily large. That is, the existence of (multiple spin) ultra-spinning MP black holes for \textit{even} $d$ is guaranteed provided that at least \textit{one} spin is exactly zero.

Consider now the case of odd $d$. With a similar arguments as those given above, one finds that a general necessary condition for having a regular horizon is $\mu>\sum^{N}_{j}\prod_{i\ne j}a_{i}^{2}$. This comes from requiring that the $r^{2}$ coefficient of $\mathcal{P}^{(2)}$ must be negative. Otherwise it would never vanish. Again, in general there will be $0$, $1$ or $2$ horizons as we proved above also for the present case of odd $d$. Now, however, the particular configuration in which the presence of an horizon is guaranteed changes with respect to the case $d$ even. Consider the spin configuration in which $m$ spins vanish with $m\geq2$. Then there is always a regular horizon regardless of the value of the rest of the spins: again writing $\mathcal{P}^{(2)}$ close to zero, we have
\begin{equation}\label{eq:54}
\mathcal{P}^{(2)}(r)\approx r^{2m}\left(\prod_{i=m+1}^{N}a_{i}^{2}\right)-\mu r^{2}
\end{equation}
which is negative for small enough $r$ if $m\geq2$ (and hence must vanish at some point since $\mathcal{P}^{(2)}$ is positive and large for large $r$), regardless of the value of the rest of the spins. For $m=1$ it is not guaranteed that $\mathcal{P}^{(2)}$ becomes negative for small enough $r$, and that will depend on the value of the rest of angular momenta. Hence, the existence of (multiple spin) ultra-spinning MP black holes for \textit{odd} $d$ is guaranteed provided that at least \textit{two} spins strictly vanish.

With this results we can proceed to obtain the phase space for the $2$-spin MP black holes in $d=5$ and $d=6$. Since these are some of the simplest multiple spin solutions, the phase spaces are easily found and they exhibit in a neat way the general behaviour predicted above about the ultra-spinning configurations.

\begin{figure}[h!]
\centering
\includegraphics[scale=0.4]{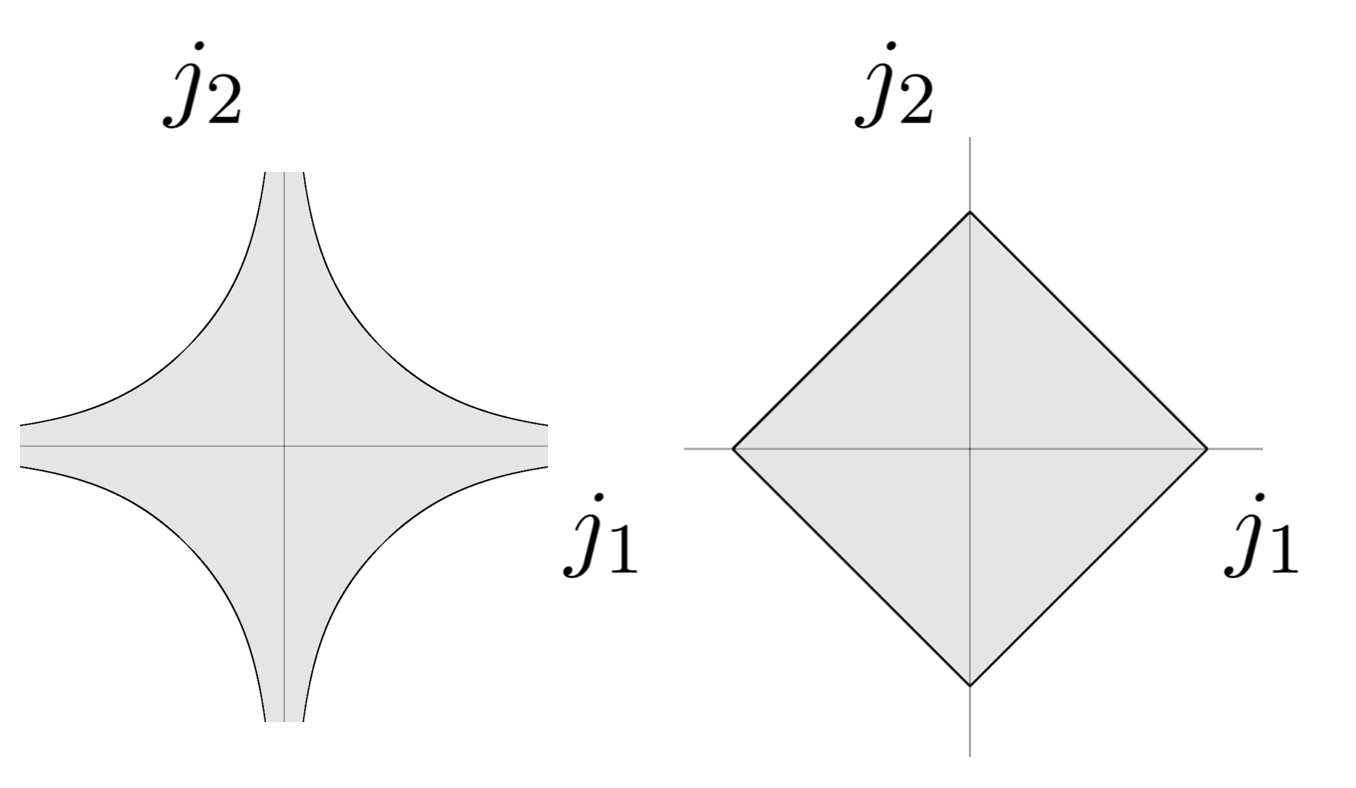}
\caption{Phase space for the 2-spin MP black hole in $d=5$ (right) and $d=6$ (left).}
\label{fig4}
\end{figure}

For $d=5$, equation $ii)$ in \eqref{eq:50} reads
\begin{equation}\label{eq:55}
r^{4}+\left(a^{2}_{1}+a^{2}_{2}-\mu\right)r^{2}+a_{1}^{2}a_{2}^{2}=0.
\end{equation}
The boundaries of the phase space in the parameter space $(j_{1},j_{2})$ will be defined by the extremal curves in which the quadratic equation \eqref{eq:55} admits only one (degenerate) positive solution for $r^{2}$. Beyond that line, no such solution for $r^{2}$ exists and, consequently, no horizon is present. Imposing a vanishing discriminant one has
\begin{equation*}
\left(\mu-\left(a_{1}^{2}+a_{2}^{2}\right)\right)^{2}-4a_{1}^{2}a_{2}^{2}=0\rightarrow \sqrt{\mu}=\lvert a_{1}\rvert+\lvert a_{2}\rvert
\end{equation*}
where we have used the condition $\mu>\sum^{N}_{j}\prod_{i\ne j}a_{i}^{2}$. Going through the respective definitions, one finds 
\begin{equation}\label{eq:56}
\lvert a_{i}\rvert=\sqrt{\frac{8GM}{3\pi}}\lvert j\rvert,\,\,\,\,\,\,\,\,\,\,\,\, \mu=\frac{8}{3\pi}GM
\end{equation}
so the extremal curves in the parameter space $(j_{1},j_{2})$ are defined by
\begin{equation}\label{eq:57}
1=\lvert j_{1}\rvert+\lvert j_{2}\rvert.
\end{equation}
This defines the contour of a diamond (see Figure \eqref{fig4}). The interior of the diamond corresponds to solutions with a non degenerate regular horizon. Along the extremal lines \eqref{eq:57} excluding the corners we have degenerate regular horizons (extremal black holes). The corners correspond to non regular horizons \cite{MPSolution,EmparanReallReview}. Notice that for $d=5$ there is no spin configuration allowing for an ultra-
spinning regime: both $j_{1}$ and $j_{2}$ are always bounded.

For the $2$-spin case in $d=6$ a bit more of work is required but the spirit is the same. The equation determining the position of the horizon reads
\begin{equation}\label{eq:58}
\Pi-\mu r=0\rightarrow r^{4}+(a^{2}_{1}+a^{2}_{2})r^{2}+a^{2}_{1}a^{2}_{2}=\mu r.
\end{equation}
The function of the LHS and its derivative are monotonically increasing polynomials, and the RHS is a straight line of slope $\mu$. These curves can intersect at, at most, two points. The extremal curve in $(j_{1},j_{2})$ of degenerate horizons is obtained by imposing that there is only one positive solution for $r$ in \eqref{eq:58}. This can be done by requiring that the curves in the LHS and RHS of \eqref{eq:58} are also tangent in the intersection point, i.e. their derivatives are equal. With this extra condition, coffee and a bit of patience, one can solve $a_{1}(\nu)$ and $a_{2}(\nu)$, being $\nu=r_{H}/\mu^{d-3}$. Then going through the respective definitions in \eqref{eq:000010}, 
\begin{equation}\label{eq:60}
\lvert j_{1}\rvert=\left(\frac{\pi}{4\sqrt{3}}\right)^{\frac{1}{3}}\sqrt{\frac{1-4\nu^{3}\pm\sqrt{1-16\nu^{3}}}{4\nu}},\,\,\,\,\,\,\, \lvert j_{2}\rvert=\left(\frac{\pi}{4\sqrt{3}}\right)^{\frac{1}{3}}\sqrt{\frac{1-4\nu^{3}\mp\sqrt{1-16\nu^{3}}}{4\nu}}
\end{equation}
where the interior root requires $0<\nu<2^{-4/3}$. As can be seen in Figure \eqref{fig4}, $j_{1}$ and $j_{2}$ can be arbitrarily large provided $j_{2}$ and $j_{1}$ are arbitrarily small, respectively. This is precisely what was predicted above for the general case of even $d$.
\subsubsection{Symmetry enhancement}
Finally, here we briefly review the isometries exhibited by the MP solutions, particularly the feature of \textit{symmetry enhancement}. The metrics \eqref{eq:45} and \eqref{eq:46} have manifest commuting isometries generated by $\partial_{t}$ and $\partial_{\phi_{i}}$. The former one-parameter sub group of isometries is isomorphic to $(\mathbb{R},+)$, while those generated by each $\partial_{\phi_{i}}$ are isomorphic to $U(1)$. Since all commute with each other they give rise to a group of isometries $\mathbb{R}\times U(1)^{N}$. For generic values of the parameters, this is the full isometry group of the spacetime up to discrete symmetries, such as the usual time reversal for rotating spacetimes $t\rightarrow-t$, $\phi_{i}\rightarrow-\phi_{i}$ for all $i$. However for special configurations of the parameters, new symmetries arise and the isometry group is therefore enhanced. Roughly speaking, switching off spins introduce rotation factors $SO$ in the isometry group, while synchronising spins (i.e. making them equal) introduces a new kind of symmetry, corresponding to combinations of rotations within planes living in the 'spin-degenerate space'.

Let us discuss first the case of vanishing spins. Take, for instance, the case in which all spins vanish but one. Then the metric becomes \eqref{eq:20} which exhibits a manifest symmetry group $\mathbb{R}\times U(1)\times SO(d-3)$. More generally, switching off $m$ spins with $d$ odd (even) leaves $2m$ ($2m+1$) coordinates in which we can define spherical coordinates and get round spherical factors of dimension $2m-1$ ($2m$) in the metric, which contain the isometry groups $SO(2m)$ ($SO(2m+1)$) \cite{isoMP}. Summing up, when $m$ spins vanish the isometry group is enhanced as
\begin{equation}\label{eq:61}
\mathbb{R}\times U(1)^{N}\rightarrow\begin{cases}\mathbb{R}\times U(1)^{(N-m)}\times SO(2m) & \text{$d$ odd}\\
\mathbb{R}\times U(1)^{(N-m)}\times SO(2m+1)&\text{$d$ even}\end{cases}.
\end{equation}
Notice that, since $SO(n)$ has as Cartan subgroup $U(1)^{\lfloor n\rfloor}$, \eqref{eq:61} is indeed an enhancement of symmetry and not a symmetry breaking process. To some extent, this is not very surprising since for all spins vanishing one recovers Schwarzschild-Tangherlini. In this sense, \eqref{eq:61} simply reflects the convergence of the MP black hole to the static, spherically symmetric case. 

Perhaps less familiar is what happens when $m$ spins synchronise. Consider \eqref{eq:45} in $d=5$ and synchronise the spins, $a_{1}=a_{2}:=a$. Pairing the spatial coordinates in planes of rotation as usual, $(x^{i},y^{i})$ with $i=1,2$, the metric can be written as
\begin{align}\label{eq:62}
ds^{2}=&-dt^{2}+\frac{(r^{2}+a^{2})}{r^{2}}\left(d{x^{1}}^{2}+d{y^{1}}^{2}+d{x^{2}}^{2}+d{y^{2}}^{2}-dr^{2}\right)+\frac{\Pi F}{\Pi-\mu r^{2}}dr^{2}+\\ \nonumber
&+\frac{\mu r^{2}}{\Pi F}\left(dt-\frac{a}{r^{2}}\left(-y^{1}dx^{1}+x^{1}dy^{1}-y^{2}dx^{2}+x^{2}dy^{2}\right)\right)^{2}.
\end{align}
The first line is manifestly symmetric under rotations within any 2-plane defined in the 'spin-degenerate' space $(x^{1},y^{1},x^{2},y^{2})$. The second line, however, is only manifestly symmetric for rotations within the original planes $(x^{1},y^{1})$ and $(x^{2},y^{2})$. A rotation within any of the $6$ orthogonal planes defined by all possible pairings of $(x^{1},y^{1},x^{2},y^{2})$ is generated by the vector field
\begin{equation}\label{eq:63}
\ell_{ab}=a\partial_{b}-b\partial_{a}
\end{equation}
where $a$ and $b$ can be any of the $x^{i}$ and $y^{i}$. As we said, \eqref{eq:62} is manifestly invariant only for $a=x^{i}$ \textit{and} $b=y^{i}$ for any $i$, but not for the more general case $\ell_{ab}$. In fact, $\ell_{ab}$ is not a Killing field in general, but with the metric written as in \eqref{eq:62} it is easy to check that combing simultaneous rotations within the 2-planes $(x^{1},x^{2})$ and $(y^{1},y^{2})$ or within $(x^{1},y^{2})$ and $(x^{2},y^{1})$ the metric is left invariant. The Killing fields generating such rotations are \cite{explicitenhancement}
\begin{equation}\label{eq:64}
\xi_{ij}=\ell_{x^{i}x^{j}}+\ell_{y^{i}y^{j}},\,\,\,\,\,\,\,\,\,\, \rho_{ij}=\ell_{x^{i}y^{j}}+\ell_{x^{j}y^{i}}.
\end{equation} 
Notice that these Killing fields also generate the original $U(1)$ symmetries, as $\frac{1}{2}\rho_{ii}$ generate the rotations within the original 2-planes. The rotations generated by \eqref{eq:64} between (and within) 2-planes form a group which is isomorphic to $U(2)$. This can be seen, for instance, defining complex coordinates $z^{1}=x^{1}+iy^{1}$, $z^{2}=x^{2}+iy^{2}$ and writing such 'rotations' as $2\times2$ complex matrices. This result is completely general \cite{isoMP}: if $m$ spins synchronise, then the original abelian symmetry factor $U(1)^{m}$ is enhanced to $U(m)$, which is non abelian. In conclusion, with $m$ spins synchronised the symmetry enhancement reads
\begin{equation}\label{eq:65}
\mathbb{R}\times U(1)^{N}\rightarrow\mathbb{R}\times U(1)^{(N-m)}\times U(m)\,\,\,\,\,\,  \text{$d$ odd and even}.
\end{equation}
Combinations of switching off some spins while synchronising others produce a symmetry enhancement that is obtained from combining \eqref{eq:61} and \eqref{eq:65} in the natural way \cite{isoMP}. It is particularly interesting the fact that, for the maximally synchronised case of $N$ spins equal, the solution becomes of cohomogeneity-1, which means that it only depends in one coordinate \cite{isoMP,explicitenhancement}.

\section{New shapes: Black Rings}
In four dimensional spacetimes, Hawking's theorem shows that the topology of cross sections of the event horizon must be $S^{2}$ \cite{Hawking,LargeScale}. Nevertheless, in $d>4$ this result does not apply and one can give heuristic arguments to understand its failure. From our work above, we already know that in higher dimensions black holes can be extended objects, and their rotation dynamics admit ultra-spinning regimes. Now take a black string $\mathcal{B}\times \mathbb{R}$ (where $\mathcal{B}$ is any vacuum black hole solution), and \textit{bend} the extended dimension of the horizon until turning it into an $S^{1}$. In this case the collapse of the ring is not stabilised by the topology as in the case of a black string $\mathcal{B}\times S^{1}$. However, the existence of ultra-spinning regimes suggest that it might be possible to balance the self-atraction with the centrifugal force due to fast rotation in the plane of the ring. That is, now the ring would be stabilised by the rotation dynamics rather than the topology. For the moment this is only a naive construction and one could think that other black hole shapes might be possible. However, the generalisation of Hawking's theorem to higher dimensions in \cite{TopologyHigherDim} still sets some constraints on the horizon topology, so not everything that we can construct with heuristic reasoning might exist.

The first exact realisation of the kind of solutions predicted above came from the work of Emparan and Reall in \cite{BROriginal}, where they obtain a \textit{black ring} in five dimensions: a stationary, asymptotically flat black hole solution whose horizon topology is $S^{1}\times S^{2}$. Furthermore, its rotation dynamics is in perfect agreement with our heuristic and naive reasoning given above. This final section is devoted to presenting such solution, focusing particularly on its rotation dynamics. The discussion on black rings closes the argumental line of this work, as they provide an example of how the combination of two phenomena, namely extended black objects and new rotation dynamics of black holes, give rise to solutions with different horizon topologies and many other features that have no four dimensional counterpart.
\subsection{Ring coordinates}
\begin{figure}[h!]
\centering
\includegraphics[scale=0.6]{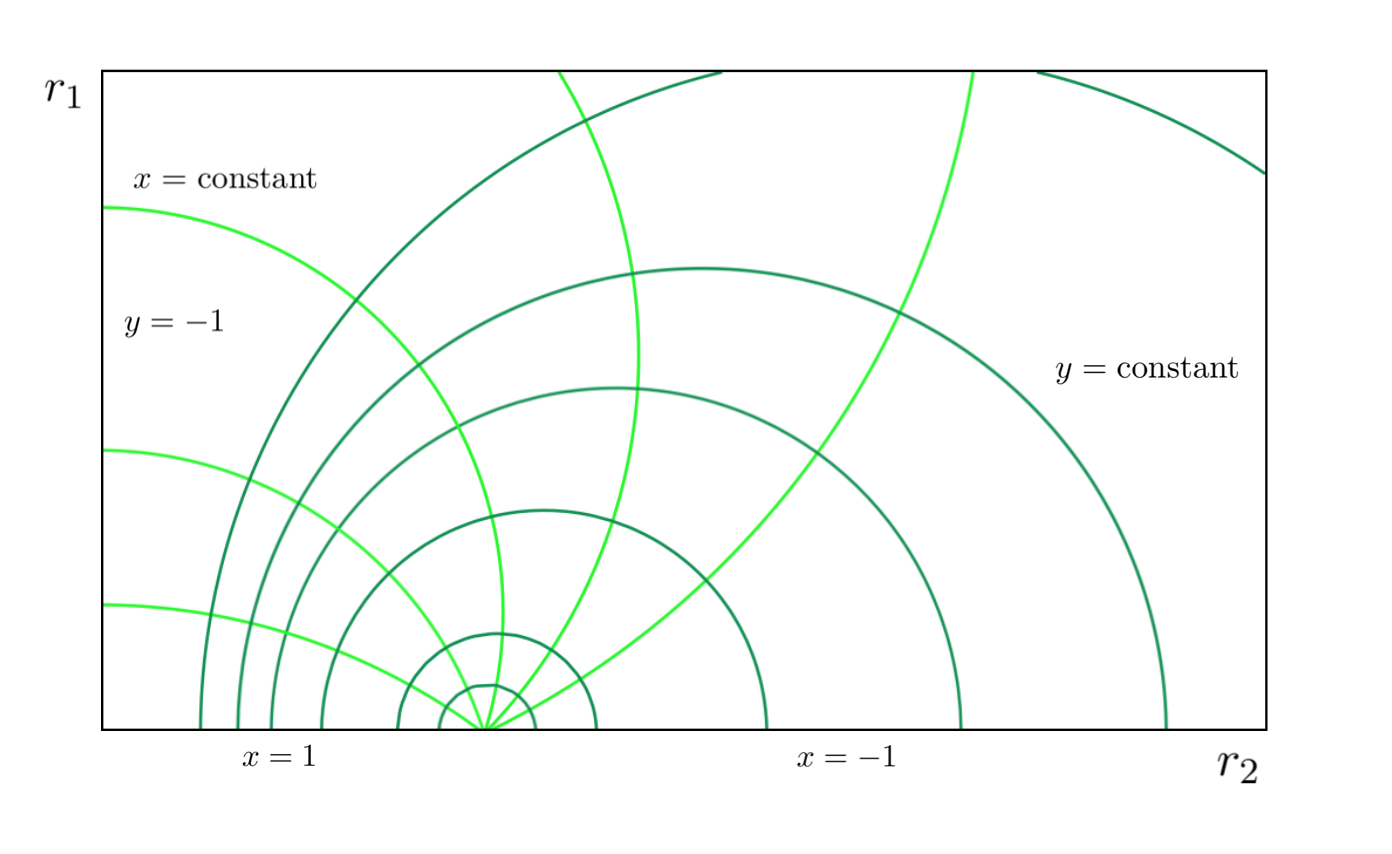}
\caption{Curves of constant $x$ (light green) and constant $y$ (dark green) in the $(r_{1},r_{2})$ plane.}
\label{fig2}
\end{figure}

A solution with horizon topology $S^{1}\times S^{2}$ is significantly different than those we have seen in four and higher dimensions so far. It is then convenient to construct new sets of coordinates that are naturally adapted to the ring and gain intuition with them. One way of doing so is by solving a problem for a field of an appropriate kind in \textit{flat space} being sourced by a distribution of charge similar to the shape of the black hole that we expect, in our case, a ring. Hence, first of all consider flat space in $d=5$ dimensions, and choose the usual coordinates $(t,r_{1},\phi,r_{2},\psi)$ in which
\begin{equation}\label{eq:111}
ds^{2}=-dt^{2}+dr_{1}^{2}+r_{1}^{2}d\phi^{2}+dr_{2}^{2}+r_{2}^{2}d\psi^{2}.
\end{equation}
Since we want coordinates adapted to a black ring, it is sensitive to choose a source placed at
\begin{equation}\label{eq:112}
r_{1}=0,\,\,\,\,\,\, r_{2}=R,\,\,\,\,\,\, 0\leq\psi<2\pi,
\end{equation}
which corresponds to a ring lying in the axis of rotation around $\phi$: the plane mapped by $(r_{2},\psi)$ at $r_{1}=0$. It turns out \cite{blackringsreview,supertubes} that for the black ring the appropriate field is a $2$-form potential $B_{\mu\nu}$ and its dual $1$-form potential $A_{\mu}$, related via $\star \text{d} B=\text{d} A$, which behave according to the equations (outside the ring source)
\begin{equation}\label{eq:A1}
\partial_{\mu}\left(\sqrt{-g}\left(\text{d}B\right)^{\mu\nu\rho}\right)=0.
\end{equation}
Given the form of the source \eqref{eq:112}, it is natural to look for solutions symmetric under the action of $\partial_{t}$, $\partial_{\phi}$ and $\partial_{\psi}$. In such case, surfaces of constant $B_{t\psi}$ will be orthogonal to surfaces of constant $A_{\phi}$. The solutions for each component $B_{t\psi}$ and $A_{\phi}$ read
\begin{align}\label{eq:114}
B_{t\psi}=-\frac{1}{2}\left(1-\frac{R^{2}+r_{1}^{2}+r_{2}^{2}}{\Sigma}\right),&\,\,\,\,\,\,\, A_{\phi}=-\frac{1}{2}\left(1+\frac{R^{2}-r_{1}^{2}-r^{2}_{2}}{\Sigma}\right),\\ \nonumber
\Sigma=&\sqrt{\left(r_{1}^{2}+r_{2}^{2}+R^{2}\right)^{2}-4R^{2}r_{2}^{2}}.
\end{align}
Now we can define coordinates $x$ and $y$ corresponding to surfaces of constant $A_{\phi}$ and constant $B_{t\psi}$, respectively, as
\begin{equation}\label{eq:116}
x=\frac{R^{2}-r_{1}^{2}-r^{2}_{2}}{\Sigma},\,\,\,\,\,\,\,\,\, y=-\frac{R^{2}+r_{1}^{2}+r_{2}^{2}}{\Sigma}
\end{equation}
with inverse
\begin{equation}\label{eq:117}
r_{1}=R\frac{\sqrt{1-x^{2}}}{x-y},\,\,\,\,\,\,\,\,r_{2}=R\frac{\sqrt{y^{2}-1}}{x-y}.
\end{equation}
Curves of constant $x$ and $y$ in the $(r_{1},r_{2})$ plane are shown in Figure\eqref{fig2}.
The metric of flat spacetime in coordinates $(t,x,y,\phi,\psi)$ reads
\begin{equation}\label{eq:118}
ds^{2}=-dt^{2}+\frac{R^{2}}{(x-y)^{2}}\left[\left(y^{2}-1\right)d\psi^{2}+\frac{dy^{2}}{y^{2}-1}+\frac{dx^{2}}{1-x^{2}}+\left(1-x^{2}\right)d\phi^{2}\right].
\end{equation} 
In order to interpret $x$ and $y$ notice the following observations: 

\begin{itemize}
\item Looking at \eqref{eq:117} we see that the ranges go as $-\infty<y\leq-1$, $-1\leq x\leq 1$.
\item We have that $g_{\psi\psi}=0$ at $y=-1$. Hence, the axis of rotation around $\psi$ lies at $y=-1$ and is mapped by $(x,\phi)$. Notice that looking at \eqref{eq:117} we see that, indeed, $y=-1$ corresponds to $r_{2}=0$. For the angular coordinate $\phi$ we have $g_{\phi\phi}=0$ (and also $r_{1}=0$) at $x=1$ and $x=-1$. Consequently, the axis of rotation around $\phi$ has two connected components and each of them is mapped by $(y,\psi)$. The component of the axis at $x=1$ has $r_{2}<R$ for all $(y,\psi)$, so it corresponds to the (open) inner disc in the plane of the ring. The component at $x=-1$, however, satisfies $r_{2}>R$ for all $(y,\psi)$ so it corresponds to the complement of the (closure of the) inner disc in the plane of the ring. 
\item Infinity lies at $y=x$ but according to the ranges of each coordinate this only happens for $y=x=-1$.
\item The source lies at $y\to-\infty$ where $r_{1}=0$ and $r_{2}=R$.
\end{itemize}

Finally, notice that moving within surfaces $y=\text{constant}$ along integral curves of $\partial_{x}$ corresponds to moving around the source, since starting at $x=1$ and finishing at $x=-1$ one goes from $r_{2}<R$ to $r_{2}>R$, respectively, without intersecting the source (see Figure\eqref{fig2}). From this one expects that the horizon of the black ring will lie at some $y=y_{H}$. As we will see in the next section, this is indeed the case.

Before presenting the black ring solution, we introduce one last set of coordinates that make manifest the horizon topology and that are particularly convenient to study the near horizon geometry of thin black rings. Define the coordinates
\begin{equation}\label{eq:119}
r=-\frac{R}{y},\,\,\,\,\,\,\,\, x=\cos\theta
\end{equation}
with ranges $0\leq r\leq R$ and $0\leq \theta\leq \pi$, in which the metric takes the form
\begin{equation}\label{eq:120}
ds^{2}=-dt^{2}+\frac{1}{\left(1+\frac{r\cos\theta}{R}\right)^{2}}\left[\left(1-\frac{r^{2}}{R^{2}}\right)R^{2}d\psi^{2}+\frac{dr^{2}}{1-\frac{r^{2}}{R^{2}}}+r^{2}\left(d\theta^{2}+\sin^{2}\theta d\phi^{2}\right)\right].
\end{equation}
Now the axis of rotation of the ring at $r_{2}=0$ corresponds to the coordinate singularity at $r=R$. Infinity is at $r=R,\theta=\pi$ and the source lies at $r=0$. The horizon will lie at some $r=r_{H}$. Finally, notice that in these coordinates the topology of (spatial cross sections of) surfaces of constant $r$ is manifestly $S^{1}\times S^{2}$. The coordinates \eqref{eq:119} will be appropriate for the near-horizon geometry of thin black rings, i.e. rings satisfying $r_{H}/R\ll1$. This is so because focusing on the region $r/R\ll1$ and expanding \eqref{eq:120} to first non trivial order, the metric reads
\begin{equation}\label{eq:A2}
ds^{2}=-dt^{2}+R^{2}d\psi^{2}+dr^{2}+r^{2}\left(d\theta^{2}+\sin^{2}\theta d\phi^{2}\right)
\end{equation}
so $r$ and $\theta$ recover their usual interpretation as the area radius and axial angle of the 2-spheres of constant $r$ and $\psi$. 
\subsection{A Black Ring with one angular momenta $J_{\psi}$}
Here we study a black ring that exhibits rotation in the plane of the ring. We focus mainly on the rotation dynamics, as well as some limit geometries that give a precise notion of the heuristic reasonings given above regarding the compensation of gravitational collapse via centrifugal force.
\subsubsection{General properties}
After the discussion on the coordinates $(x,y)$ above, we are now ready to read properly the black ring solution in one of its most convenient forms \cite{BRDynamics},
\begin{align}\label{eq:123}
ds^{2}=&-\frac{F(y)}{F(x)}\left(dt-CR\frac{1+y}{F(y)}d\psi\right)^{2}+\\ \nonumber
&+\frac{R^{2}}{(x-y)^{2}}F(x)\left[-\frac{G(y)}{F(y)}d\psi^{2}-\frac{dy^{2}}{G(y)}+\frac{dx^{2}}{G(x)}+\frac{G(x)}{F(x)}d\phi^{2}\right].
\end{align}
where
\begin{align}\label{eq:124}
F(\xi)=1+\lambda \xi,&\,\,\,\,\,\,\, G(\xi)=(1-\xi^{2})\left(1+\nu\xi\right),\\ \nonumber
C=&\sqrt{\lambda(\lambda-\nu)\frac{1+\lambda}{1-\lambda}}
\end{align}
and to guarantee that $C\in\mathbb{R}$ the range of the dimensionless parameters $\nu,\lambda$ must satisfy
\begin{equation}\label{eq:124a}
0<\nu\leq\lambda<1.
\end{equation}
Switching off $\lambda$ and $\nu$, this solution becomes exactly \eqref{eq:118}. Consequently, we can regard \eqref{eq:123} just as \eqref{eq:118} but allowing $F(\xi)$ and $G(\xi)$ to depart from their flat form (which is recovered for $\lambda=\nu=0$) due to the presence of curvature. Such curvature approaches zero as $x,y\to-1$ which corresponds to infinity. The ranges of the coordinates $x$ and $y$ that are connected to infinity and in which the metric \eqref{eq:123} is regular are
\begin{equation}\label{eq:A3}
-1<x<1,\,\,\,\,\,\, -1/\nu<y<-1.
\end{equation}
In what follows we will see that $x=\pm 1$ and $y=-1$ correspond to either coordinate or conical singularities, while $y=-1/\nu$ is a mere coordinate singularity corresponding to the event horizon\footnote{In this sense, requiring $-1/\nu<y<-1$ is analogous to requiring $2M<r<\infty$ in the Schwarzschild black hole when written in Schwarzschild coordinates.}. The presence of the term $\sim dtd\psi$ in \eqref{eq:123} indicates rotation around $\psi$. There is another immediate observation about this solution, based again on primitive physical intuition: for a given mass and radius of the black ring, one expects the dynamics to fix the angular momenta required to compensate the gravitational collapse. According to this, the black ring \eqref{eq:123} should belong to a $2$-parameter family of solutions rather than living in a $3$-parameter space $(R,\lambda,\nu)$. A slightly more accurate inspection of \eqref{eq:123} gives a beautiful solution to this apparent contradiction with our physical intuition. The answer lies in the periods of the angular coordinates $\phi$ and $\psi$, which have not been specified so far. Let us see how this works. The fact that $G(\xi)$ vanishes both at $\xi=1$ and $\xi=-1$ indicates the presence of an axis of rotation around $\psi$ at $y=-1$, and around $\phi$ with two connected components one at $x=-1$ and the other at $x=1$. Following an analogous reasoning to that performed for flat space above, we learn that the axis at $y=-1$ is the axis of rotation of the ring, while $x=1$ and $x=-1$ correspond respectively to the positions of the inner and outer components (meaning $r_{2}<R$ and $r_{2}>R$ for $r_{2}$ a function given by \eqref{eq:117}) of the axis of rotation of the 2-spheres of constant $r$ (as defined in \eqref{eq:119}) and $\psi$. Let us see if there are angular defects or excesses \textit{along} each axis. Writing the metric \eqref{eq:123} close to the (spatial sections of the) axis of the ring lying at $y=-1$, one has
\begin{equation}\label{eq:124b}
ds^{2}=\frac{R^{2}}{(x+1)^{2}}F(x)\left[-\frac{G'(-1)(y+1)}{F(-1)}d\psi^{2}-\frac{dy^{2}}{G'(-1)(y+1)}+\frac{dx^{2}}{G(x)}+\frac{G(x)}{F(x)}d\phi^{2}\right].
\end{equation} 
Performing a coordinate transformation\footnote{I am grateful to my supervisor Dr.Santos for clarifications on this point.}
\begin{equation}\label{eq:124d}
y=-1+\alpha \rho^{\beta}\,\,\,\,\,\,\,\,\, \alpha,\beta\in\mathbb{R}
\end{equation}
and choosing $\alpha=-G'(-1)/4$ and $\beta=2$, the metric becomes
\begin{equation}\label{eq:124e}
ds^{2}=\frac{R^{2}}{(x+1)^{2}}F(x)\left[\frac{G'(-1)^{2}}{4F(-1)}\rho^{2}d\psi^{2}+d\rho^{2}+\frac{dx^{2}}{G(x)}+\frac{G(x)}{F(x)}d\phi^{2}\right].
\end{equation}
At each point along the axis the last two terms in \eqref{eq:124e} are regular (with the exceptions of $x=\pm 1$ that we will consider later), and the first two are conformal to $\mathbb{E}^{2}$ if the angular coordinate
\begin{equation}\label{eq:124f}
\tilde{\psi}=\frac{\lvert G'(-1)\rvert}{2\sqrt{F(-1)}}\psi=\frac{1-\nu}{\sqrt{1-\lambda}}\psi
\end{equation}
is canonically identified, that is, $\tilde{\psi}\sim\tilde{\psi}+2\pi$. In turn, this means that the original ring angle $\psi$ must have period
\begin{equation}\label{eq:124g}
\psi\sim\psi+2\pi\frac{\sqrt{1-\lambda}}{1-\nu}
\end{equation}
in order to avoid conical singularities. The analogue analysis for the outer component of the axis of the 2-spheres, lying at $x=-1$, gives that with the same identification for $\phi$,
\begin{equation}\label{eq:124h}
\phi\sim \phi+2\pi\frac{\sqrt{1-\lambda}}{1-\nu},
\end{equation}
there are no angular defects/excesses along such axis. Notice that imposing simultaneously \eqref{eq:124g} and \eqref{eq:124h} implies that no conical singularities are present at infinity, which lies at $x=y=-1$. However, when considering the inner axis at $x=1$ one finds that 
\begin{equation}\label{eq:124j}
\tilde{\phi}=\frac{\lvert G'(1)\rvert}{2\sqrt{F(1)}}\phi=\frac{1+\nu}{\sqrt{1+\lambda}}\phi
\end{equation}
must be canonically identified to avoid angular defects/excesses along such axis. The corresponding period for $\phi$ is
\begin{equation}\label{eq:124k}
\phi\sim\phi+2\pi\frac{\sqrt{1+\lambda}}{1+\nu}
\end{equation}
which for generic values of $\lambda$ and $\nu$ is different from the period in \eqref{eq:124h}. Only if
\begin{equation}\label{eq:124l}
\lambda=\frac{2\nu}{1+\nu^{2}}
\end{equation}
will those periods coincide. Summing up, no conical singularities will be present along any of the axes (including their intersection points at the centre of the ring and at infinity) if \eqref{eq:124l} holds and $(\psi,\phi)$ are identified as in \eqref{eq:124g} and \eqref{eq:124h}, respectively. If then we drop \eqref{eq:124l}, we will have regularity at infinity, but a conical singularity will be present along the inner component of the axis of the 2-spheres. The constraint \eqref{eq:124l} eliminates one parameter in \eqref{eq:123} making the solution fully specified with only two parameters, according to our initial intuition. As we will see in the following sections, conical singularities correspond to the presence of certain tensions along the ring: an angular defect makes the ring "shorter" so a tension is required to make static such configuration. Antagonically, an angular excess makes the ring "larger" so some pressure (negative tension) is required to keep the ring in that larger shape. Not by chance, the reader might find in this certain reminiscence of Newtonian mechanics. We will make this resemblance precise in the following sections.

Let us now discuss the pathologies in the metric components of \eqref{eq:123} \footnote{In this study of general properties, we are only going to compute explicitly the magnitudes related to the rotation of the ring, as these are the ones we will use later. For the other results that we only cite here, we will refer the reader essentially to \cite{BROriginal} and \cite{blackringsreview}}. In what follows, we do not impose \eqref{eq:124l}, unless stated otherwise. At $y=-1/\lambda$ the function $F(y)$ vanishes, but both the metric and its inverse are smooth \cite{blackringsreview}. Actually, a more careful observation reveals that
\begin{equation}\label{eq:135}
\left(\partial_{t}\right)^{2}=-\frac{F(y)}{F(x)}\xrightarrow[y\to-\frac{1}{\lambda}]{}0.
\end{equation}
That is, $y=-1/\lambda$ is an ergosurface: stationary observers following integral curves of $\partial_{t}$ become space like when crossing $y=-1/\lambda$ inwards. At $y=-1/\nu$, $G(y)$ vanishes so $g_{yy}$ diverges. Nevertheless, following the usual procedure to define the (analogue of the) Eddington-Finkelstein coordinates in which $\partial_{y}$ is null, 
\begin{equation}\label{eq:A4}
\text{d}t=\text{d}v-CR\frac{1+y}{G(y)\sqrt{-F(y)}}\text{d}y,\,\,\,\,\,\, \text{d}\psi=\text{d}\psi'+\frac{\sqrt{-F(y)}}{G(y)}\text{d}y
\end{equation}
the metric becomes 
\begin{equation}\label{eq:A5}
ds^{2}=-\frac{F(y)}{F(x)}\left(dv-CR\frac{1+y}{F(y)}d\psi'\right)^{2}+\frac{R^{2}}{(x-y)^{2}}F(x)\left[-\frac{G(y)}{F(y)}d\psi'^{2}+2\frac{d\psi'dy}{\sqrt{-F(y)}}+\frac{dx^{2}}{G(x)}+\frac{G(x)}{F(x)}d\phi^{2}\right]
\end{equation}
which is regular at $y=-1/\nu$ \cite{BROriginal}. Indeed, $\text{d}y$ becomes null as $y\to-1/\nu$, so that the hyper surfaces of $y=\text{constant}$ become null at $y=-1/\nu$. This, in turn, indicates that $y=-1/\nu$ is the event horizon of the black ring \cite{BROriginal}. Such horizon hides a curvature singularity at $y\to-\infty$ where $R^{\mu\nu\sigma\rho}R_{\mu\nu\sigma\rho}$ diverges \cite{blackringsreview}. The event horizon is also a Killing horizon, and as one would expect for a rotating black hole, its normal is not the Killing field defining stationarity at infinity, $\partial_{t}$. In order to obtain the Killing field for which $y=-1/\nu$ is a Killing horizon, we can first compute the angular velocity of the black hole. This can be easily obtained by computing the angular velocity\footnote{The angular velocity is wrt stationary observers at infinity, i.e. $d\tilde{\psi}/dt=u^{\tilde{\psi}}/u^{t}$.}of the observers "locally at rest", with tangent $u^{a}=dt^{a}/\sqrt{\lvert dt_{b}dt^{b}\rvert}$, close to the horizon (see, e.g. \cite{WaldGR}). 
The angular velocity of the horizon is then given by
\begin{equation}\label{eq:139}
\Omega_{H}=\lim_{y\to y_{h}}\frac{u^{\tilde{\psi}}}{u^{t}}=\lim_{y\to y_{h}}-\frac{g_{t\tilde{\psi}}}{g_{\tilde{\psi}\tilde{\psi}}}=\frac{1}{R}\sqrt{\frac{\lambda-\nu}{\lambda(\lambda+1)}}
\end{equation}
where in the second equality we have used that $u\cdot\partial_{\tilde{\psi}}=0$. Indeed, trying with the Killing field
\begin{equation}
V=\partial_{t}+\Omega_{H}\partial_{\tilde{\psi}}
\end{equation}
one can check via standard calculations that on the horizon $V_{a}$ is proportional to $\left(\text{d}y\right)_{a}$ \cite{BROriginal}. This confirms that \eqref{eq:139} is the angular velocity of the black hole.

For completeness of this section and preparing what will be done in the next ones, now we shall present the black ring in the more intuitive coordinates $(r,\theta)$, as defined in \eqref{eq:119}. First, in order to interpret $\nu$ and $\lambda$ it is convenient to rewrite them in terms of two new parameters $(r_{H},\sigma)$,  
\begin{equation}\label{eq:140}
\nu=\frac{r_{H}}{R},\,\,\,\,\,\,\,\,\, \lambda=\frac{r_{H}\cosh^{2}\sigma}{R}.
\end{equation}
Looking at \eqref{eq:119}, we see that $r_{H}=R\nu$ is precisely the position of the horizon. From this we learn that $\nu$ is a shape parameter measuring (approximately) the ratio between the radius of the 2-spheres of the horizon and the radius of the ring. The interpretation of $\sigma$ will become clear in the next sections. In the chart $(r,\theta)$ and in terms of the new parameters $(r_{H},\sigma)$, the metric reads \cite{blackringsreview}
\begin{align}\label{eq:141}
ds^{2}=&-\frac{\hat{f}}{\hat{g}}\left(dt-r_{H}\sinh\sigma\cosh\sigma\sqrt{\frac{R+r_{H}\cosh^{2}\sigma}{R-r_{H}\cosh^{2}\sigma}}\frac{\frac{r}{R}-1}{r\hat{f}}Rd\psi\right)^{2}+\\ \nonumber
&+\frac{\hat{g}}{\left(1+\frac{r\cos\theta}{R}\right)^{2}}\left[\frac{f}{\hat{f}}\left(1-\frac{r^{2}}{R^{2}}\right)R^{2}d\psi^{2}+\frac{dr^{2}}{\left(1-\frac{r^{2}}{R^{2}}\right)f}+\frac{r^{2}}{g}d\theta^{2}+\frac{g}{\hat{g}}r^{2}\sin^{2}\theta d\phi^{2}\right]
\end{align}
with
\begin{equation}\label{eq:142}
f=1-\frac{r_{H}}{r},\,\,\,\,\,\,\,\, \hat{f}=1-\frac{r_{H}\cosh^{2}\sigma}{r}
\end{equation}
and 
\begin{equation}\label{eq:143}
g=1+\frac{r_{H}}{R}\cos\theta,\,\,\,\,\,\,\,\, \hat{g}=1+\frac{r_{H}\cosh^{2}\sigma}{R}\cos\theta.
\end{equation}
As a final remark, notice that in this chart both the horizon at $r=r_{H}$ and the ergosurface at $r=r_{H}\cosh^{2}\sigma$ exhibit manifest ring topology $S^{1}\times S^{2}$.
\subsection{Black Hole non-uniqueness in $d=5$}
The family of solutions \eqref{eq:123} breaks explicitly black hole uniqueness in $d=5$. That is, it constitutes an explicit example that stationary $U(1)^{2}$-symmetric vacuum black holes in $d=5$ are \textit{not} uniquely specified by their mass and angular momenta. This was first noticed by the same authors who obtained the black ring solution \cite{BROriginal}, and constitutes one of the main consequences of their discovery. This added to the existence of MP black holes gives a triple degeneracy of solutions in certain region of the (one-dimensional) phase space of single spinning black holes in five dimensions. A neat way of seeing this is by comparing the relation between the horizon area and the angular momenta of both the black ring and the single spinning MP black hole. This comparison was shown for the first time in \cite{EndBHnonuniqueness}. 

\begin{figure}[h!]
\centering
\includegraphics[scale=0.4]{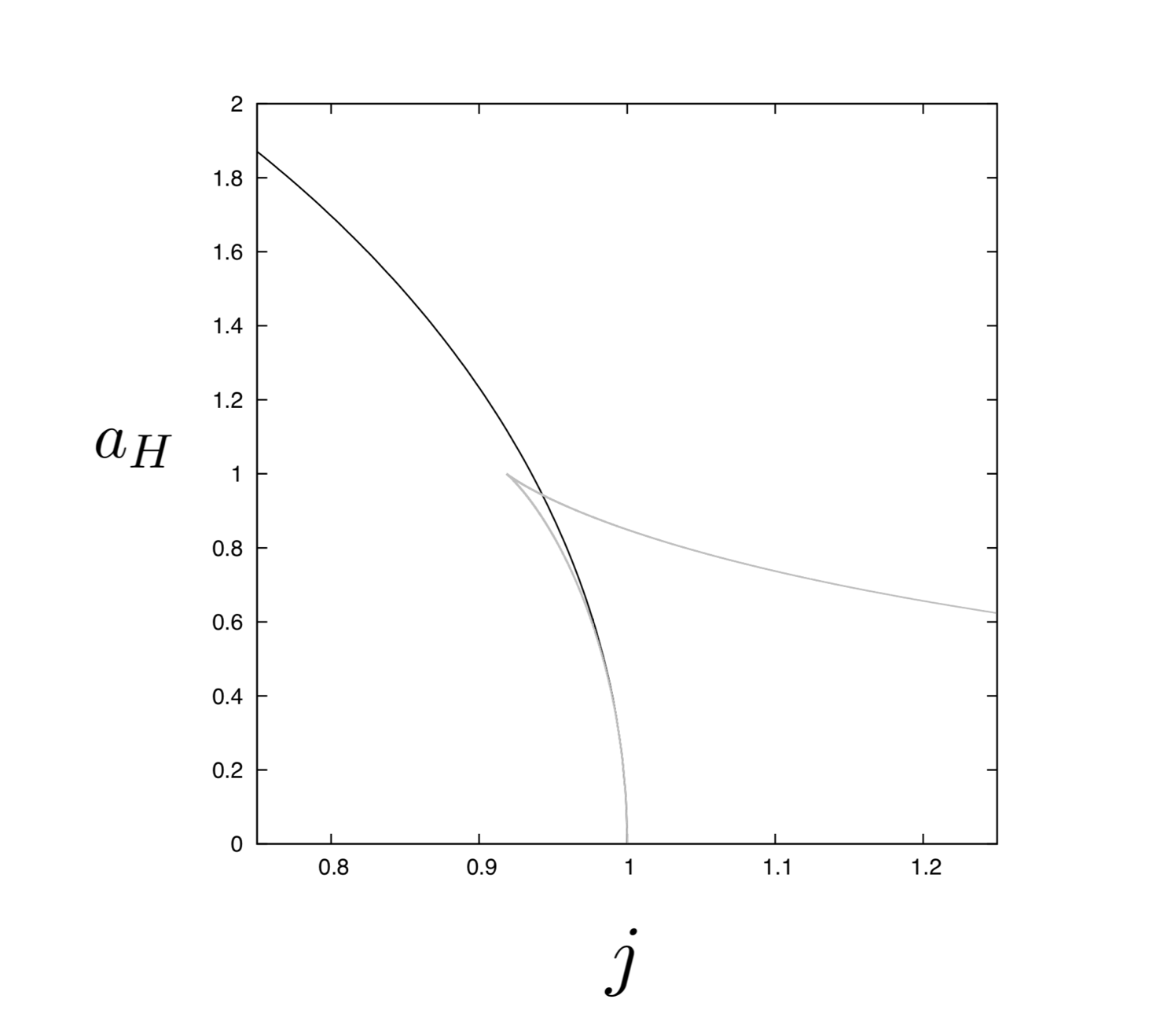}
\caption{Curve $a_{H}(j)$ for MP (black) and black rings (gray).}
\label{fig3}
\end{figure}
In the previous section we obtained the curve $a_{H}(j)$ for the five dimensional, single spinning MP black hole. In a similar way one can obtain the corresponding curve for black rings. First, by writing the solution \eqref{eq:123} in terms of the angles $(\tilde{\psi},\tilde{\phi})$ and studying the asymptotic field one identifies the mass and the angular momenta just as how we did in the MP black hole \cite{blackringsreview},
\begin{equation}\label{eq:143a}
M=\frac{3\pi R^{2}}{4G}\frac{\lambda}{1-\nu},\,\,\,\,\,\, J=\frac{\pi R^{3}}{2G}\frac{\sqrt{\lambda(\lambda-\nu)(\lambda+1)}}{(1-\nu)^{2}}.
\end{equation}
Computing the area of (spatial cross sections of) surfaces of constant $y$ in \eqref{eq:123} and making $y\to-1/\nu$, one gets
\begin{equation}\label{eq:143b}
\mathcal{A}_{H}=8\pi^{2}R^{3}\frac{\nu^{3/2}\sqrt{\lambda(1-\lambda^{2})}}{(1-\nu)^{2}(1+\nu)}.
\end{equation}
All these magnitudes depend on both parameters $\lambda$ and $\nu$, but in the previous section we found that at equilibrium (i.e. in absence of external tensions along the ring) such parameters are related via \eqref{eq:124l}. We are interested in studying solutions at equilibrium so we can use such relation to eliminate one of the parameters. Since $\nu$ has a clear interpretation as a shape parameter controlling the thickness of the ring, the convenient choice is to eliminate $\lambda$. Then, using the mass to fix the scale and going through the definitions one obtains the curve $a_{H}(j)$ for the black ring in parametric form
\begin{equation}\label{eq:143c}
a_{H}=2\sqrt{\nu(1-\nu)},\,\,\,\,\,\,\,\,\, j=\sqrt{\frac{(1+\nu)^{3}}{8\nu}},
\end{equation}
where recall that, from \eqref{eq:124a}, $0<\nu<1$. A comparison of $a_{H}(j)$ for the black ring and the MP black hole is shown in Figure\eqref{fig3}. Notice the following three main observations: first, the cusp in Figure\eqref{fig3} of the black ring curve indicates the existence of two families of solutions, a thin ring family corresponding to the range $0<\nu<1/2$, and a plump ring family lying in $1/2<\nu<1$. Secondly, between $j=\sqrt{27/32}$ and $j=1$, i.e. in the range of existence of the plump family of rings, we have a triple degeneracy of black hole solutions. In other words, in that region of $j$ black hole uniqueness in $d=5$ is broken explicitly. Since the topology $S^{1}\times S^{2}$ is also degenerated in $\sqrt{27/32}<j<1$, we learn that the horizon topology does not determine uniquely black holes in $d=5$ either. Finally, notice that the plump branch meets the MP curve in the extremal point $(j=1,a_{H}=0)$. This suggests that black rings and MP black holes might be connected in some way. We will come back to this point in the following sections. 
\subsection{Black Rings from Black Strings}
At the very beginning of this chapter, we gave some naive arguments that pointed towards the existence of black holes with the topology of a ring in $d>4$. The core of the reasoning is the heuristic idea that extended black objects with 'contractible' dimensions should be able to compensate their collapse by rotating fast enough, as we have seen that in $d>4$ rotation is not necessarily limited. The purpose of this section is to give a precise notion of such idea.

Consider a thin black ring (i.e., a ring for which the radius of the $S^{1}$ is much larger than that of the $S^{2}$) which is rotating in the plane of the ring. Black rings of this kind live in the sector $\nu\ll1$ of our solution \eqref{eq:123}. Now consider a region which is close to the horizon, $r\approx r_{H}$. Since the ring is rotating, and the diametrically opposite points are far from each other one expects that an observer in that region would perceive the field created by a \textit{boosted black string}. Let us write the metric of such object. Consider a \textit{straight} black string, constructed by performing the product of Schwarzschild-Tangherlini in $d-1$ dimensions and the flat real line $\mathbb{R}$ with coordinate $z'$. Now perform a Lorentz boost\footnote{Recall that the velocity of the boost $v$ and the rapidity $\sigma$ are related by $v=\tanh{\sigma}$.} in the direction of the string,
\begin{equation}\label{eq:145}
t'=t\cosh{\sigma}-z\sinh{\sigma},\,\,\,\,\,\,\, z'=z\cosh{\sigma}-t\sinh{\sigma}.
\end{equation} 
The metric becomes
\begin{align}\label{eq:146}
ds^{2}=&-\left(1-\cosh^{2}\sigma\frac{r_{H}^{d-4}}{r^{d-4}}\right)dt^{2}-2\frac{r_{H}^{d-4}}{r^{d-4}}\cosh\sigma\sinh\sigma dtdz+\left(1+\sinh^{2}\sigma\frac{r_{H}^{d-4}}{r^{d-4}}\right)dz^{2}+\\ \nonumber
&+\frac{dr^{2}}{\left(1-\frac{r_{H}^{d-4}}{r^{d-4}}\right)}+r^{2}d\Omega_{d-3}.
\end{align}
A boosted \textit{circular} black string of radius $R$ is obtained by identifying
\begin{equation}\label{eq:147}
z\sim z+2\pi R.
\end{equation}
In that case it is convenient to introduce a new angular coordinate $\psi=z/R$ which is canonically identified. Following our reasoning above, the black ring \eqref{eq:141} should become \eqref{eq:146} in the region around the horizon $r\approx r_{H}$ in the limit of a thin ring $r_{H}\ll R$. Indeed, by inspection of \eqref{eq:141} one can see that in the limit \cite{blackringsreview,BRBlackStringLimit}
\begin{equation}\label{eq:148}
r,r_{H},r_{H}\cosh^{2}\sigma\ll R,
\end{equation}
the metric \eqref{eq:146} is recovered for $d=5$ after defining $z=R\psi$ (with appropriate identification to avoid angular defects). In other words, in the limit \eqref{eq:148} the black ring becomes a circular boosted black string with radius $R$. Such limit corresponds to $i)$ making the black ring very thin as $r_{H}/R\ll 1$ and $ii)$ focussing in a region around the ring that is small compared to its radius, $r\ll R$ (which gives a notion of staying close to the horizon wrt \textit{R}, $r\approx r_{H}$). This is exactly the expected result, and gives a precise notion of the heuristic construction of a black ring that we performed above: we can regard a black ring as a boosted circular black string, in the sense that \eqref{eq:123} contains the limit geometry \eqref{eq:146}. At this point it is interesting to mention that this kind of reasoning provides a powerful tool to solve Einstein's equations approximately. A black ring in $d\geq6$ has still not been found (mainly due to the lack of systematic solution-generating techniques in that number of dimensions, \cite{EmparanReallReview}). However, one can try to obtain an approximate solution by matching the expected near horizon geometry of thin rings \eqref{eq:146} to their expected asymptotic field. This procedure, generally known as \textit{matched asymptotic expansion}, was originally used to obtain approximate black hole solutions localised around a point of a compact extra dimension (typically $S^{1}$, motivated by the Kaluza-Klein theory) \cite{MatchedAsym1,MatchedAsym2,MatchedAsym3}. It is in \cite{BRMatchedAsym} that this procedure was adapted to the case of thin black rings with a single angular momenta, corresponding to rotation around the plane of the ring. The near horizon geometry proposed by the authors is precisely that of a boosted circular black string \eqref{eq:146}, while the asymptotic weak field solution is chosen to be that created by a circular distribution of a given mass and momentum density.

Moving back to our study of the thin black ring limit, now we see that the interpretation of $\sigma$ in \eqref{eq:140} is manifest: it plays the role of the boost parameter (or rapidity) of the circular black string that approximates the near horizon geometry of thin black rings. Let us close this section completing the interpretation of the limit in \eqref{eq:148}. The last requirement that remains to be interpreted is
\begin{equation}\label{eq:150}
\frac{R/r_{H}}{\cosh^{2}\sigma}\gg1.
\end{equation}
Consider a straight black string with horizon at $r_{H}$ that is boosted close to the speed of light, i.e. with very large $\lvert \sigma\rvert$ (and consequently very large $\cosh^{2}\sigma$). Then \eqref{eq:150} tells us that a black ring rotating very fast resembles a circular boosted black string if it is thin enough. This has a clear heuristic interpretation: a rotating object produces geometrical effects such as the dragging of inertial frames, and the faster the rotation is, the more significant these effects become. A fast rotating black ring will only resemble a boosted black string in a region close to the horizon if the diametrically opposed region is far enough to make negligible its rotation effects on the geometry of the former region.

All this discussion seems to reveal a close correspondence between our most basic, Newtonian intuition, and the mechanics of thin rings. One can wonder up to what point is this correspondence precise. The answer is that the resemblance between the mechanics of thin black rings and Newtonian strings is \textit{exact} in some cases. The precise formulation of this rather surprising statement is given in the following section.

\subsection{Newtonian Strings and thin Black Rings}
In order to study the dynamics of black rings, first we have to define a precise notion of the tension arising from the presence of conical singularities. Henceforth, assume the identification periods in \eqref{eq:124g} and \eqref{eq:124h}, which guarantee asymptotical flatness, but drop the equilibrium condition \eqref{eq:124l}. From the discussion above, we know that in such case an angular defect/excess will be present along the inner component of the axis of rotation of the 2-spheres. It is given by
\begin{equation}\label{eq:151}
\delta=2\pi-\frac{1+\nu}{\sqrt{1+\lambda}}\Delta\phi=2\pi\left(1-\frac{1+\nu}{1-\nu}\sqrt{\frac{1-\lambda}{1+\lambda}}\right).
\end{equation}
where $\Delta\phi$ is the period in \eqref{eq:124h}. The interpretation of this is clear: for generic values of $\lambda$ and $\nu$ the $2$-spheres of the horizon suffer from an azimuthal angular defect/excess with respect to the equilibrium configuration. Consequently, an external tension/pressure along the $S^{1}$ of the ring is needed in order to keep static the horizon with the squeezed/dilated 2-spheres, respectively. Regarding the pressure as negative tension, we follow \cite{BRDynamics} and define a force per unit length along the ring as\footnote{The choice of the numerical factor in front of $\delta/G$ is motivated by the correspondence to Newtonian mechanics that will become clear soon.}
\begin{equation}\label{eq:152}
\tau=\frac{2}{16\pi G}\delta=\frac{3}{8G}\left(1-\frac{1+\nu}{1-\nu}\sqrt{\frac{1-\lambda}{1+\lambda}}\right).
\end{equation}
If there is an angular defect then $\delta>0$, so there is indeed an external tension $\tau>0$. On the other hand, an angular excess requires some external pressure $\tau<0$ to keep the system static. Finally, if there are no angular pathologies then $\delta=0$ and no external forces are needed to keep staticity, as one would expect at equilibrium. In other words, \eqref{eq:152} gives a notion of external tension according to our initial intuition. The objective now is to understand the dynamics of the thin black rings, i.e. obtain the relation between the dynamical variables. It is reasonable to expect that the relevant magnitudes for the ring dynamics are: its proper radius $R_{1}$, mass $M$, angular velocity $\Omega_{H}$ and the external tension $\tau$. Whereas for general black rings there is no unique notion of proper radius $R_{1}$ (one can compute different proper radius along different choices of $\theta$), in the case of thin rings we can define the proper radius uniquely as the proper radius of the corresponding circular boosted black string. Explicitly, the proper length of a curve $\gamma(z)=(t_{0},r_{H},\theta_{0},z,\phi_{0})$, $0<z<R$, in the limit geometry \eqref{eq:146} gives
\begin{equation}\label{eq:153}
R_{1}=\frac{1}{2\pi}\int_{0}^{R}dz\sqrt{\lvert g_{zz}\rvert}=\sqrt{1+\sinh^{2}\sigma}R=\cosh\sigma R=\sqrt{\frac{\lambda}{\nu}}R.
\end{equation}
The rest of dynamical magnitudes in the thin black ring limit \eqref{eq:148}, in which both $\nu$ and $\lambda$ are very small, become
\begin{equation}\label{eq:154}
M=\frac{3\pi R^{2}}{4G}\lambda,\,\,\,\,\,\, \Omega_{H}=\frac{1}{R}\sqrt{1-\frac{\nu}{\lambda}}, \,\,\,\,\,\, \tau=\frac{3}{8G}(\lambda-2\nu). 
\end{equation}
Elimination of the unphysical parameters $\lambda$, $\nu$ and $R$ using the equations \eqref{eq:154} gives
\begin{equation}\label{eq:155}
M\Omega_{H}^{2}R_{1}=2\pi R_{1}\tau+\frac{M}{R_{1}}.
\end{equation}
It is natural to identify the first term in the RHS as the total force exerted along the ring by the external tension $\tau$, $F_{\tau}=2\pi R_{1}\tau$. It is also natural to identify the second term with a string tension, defined as the energy per unit length $\mathcal{T}=M/2\pi R_{1}$. We can now write the law of thin black ring mechanics as
\begin{equation}\label{eq:156}
M\Omega_{H}^{2}R_{1}=F_{\tau}+2\pi\mathcal{T}.
\end{equation}
One immediately realises that this is \textit{exactly} the same law that rules the mechanics of Newtonian strings. This gives a reason of why our initial heuristic arguments were in so good agreement with the results obtained for thin black rings. The result \eqref{eq:156} was first obtained in \cite{BRDynamics}, together with the general study of black ring mechanics.
\subsection{Connecting Black Holes to Black Rings}
When studying the curves $a_{H}(j)$ for the MP black hole and the black ring, we found that the plump branch converges to the MP curve as $j$ goes to 1 from below. This can be regarded as an argument in favour of a possible connection between the MP and black ring solutions, in the sense that one solution is recovered from certain limit of the other. There are also other arguments in favour of such connection. Computing the inner and outer radius of the horizon in the plane of the ring\footnote{That is, $1/2\pi$ times the proper length of closed loops along $\partial_{\psi}$ within the horizon at $x=1$ and $x=-1$, respectively.} one finds  
\begin{equation}\label{eq:157}
R_{o}=R\sqrt{\frac{\lambda(1+\lambda)}{\nu(1-\lambda)}},\,\,\,\,\,\, R_{i}=R\sqrt{\frac{\lambda}{\nu}}.
\end{equation}
When the black ring approaches the end point of the plump family as $\nu\to1$ (and hence $\lambda\to1$), we have that the outer radius diverges $R_{o}\to\infty$ while the interior radius remains finite $R_{i}\to R$. This is reminiscent of what we obtained above for the extremal MP black hole in $d=5$: in that case the confinement radius also diverges when approaching the Kerr bound. If in addition of $\nu\to1$ we require $R\to0$ in an appropriate way, the interior radius vanishes while the exterior one diverges. The ring would be spread along the whole rotation plane, just as what happens with the extremal MP black hole in $d=5$.

Indeed, taking the limit $\nu\to1$, $\lambda\to1$ and $R\to0$ while keeping the angular momenta and mass in \eqref{eq:143a} finite, one recovers the single spin MP solution from \eqref{eq:123}. The coordinate transformation that, in such limit, brings \eqref{eq:123} to the form \eqref{eq:20} is given in \cite{blackringsreview}.

\section{Final Comment}
While it might be possible that in $d=5$ we already know all stationary, asymptotically flat solutions with two abelian $U(1)$ symmetries, the absence of solution generating techniques in $d\geq6$ make such sector of the theory \textit{terra incognita}, and only approximate calculations, numerical results and heuristic arguments have provided some insight. Furthermore, the stability and endpoints of currently known solutions in $d>4$ is still a widely open field, with a very intense research activity. One of the main reasons for the interest in evolving numerically perturbed solutions in $d>4$ is that they have been seen to violate the weak cosmic censorship conjecture (see, for example, \cite{Pau} for reasonably recent work). This goes in the line that, perhaps, in order to understand why gravity is like it is in $d=4$, one might study what is different when $d>4$. For these reasons and several others, GR in higher dimensions constitutes a topic of great interest within the Scientific Community with a large amount of open questions that will certainly provide exciting research in the following years.

\vspace{1.5cm}

\large\textsc{Acknowledgements}.
\\
I am grateful to my supervisor Dr. Jorge Santos for useful guidance and discussion. I would also like to sincerely thank my Part III colleague Igor Arrieta and Biel Cardona for always being kind to discuss various issues with me. During my stay in Cambridge I have been supported by "La Caixa" fellowship.


\end{document}